\pdfoutput=1
\documentclass[USenglish,oneside,twocolumn]{article}
\usepackage[big]{dgruyter_new}

\usepackage{amsmath,amssymb,amsfonts}
\usepackage{bm}
\usepackage{algorithmic}
\usepackage{graphicx}
\usepackage{textcomp}
\usepackage{xcolor}
\usepackage{enumitem}
\usepackage[labelfont=bf]{caption}
\usepackage[T1]{fontenc}
\usepackage{mathtools}
\usepackage{subfigure}
\usepackage{comment,wasysym}
\usepackage{color, soul}
\usepackage{pbox}
\usepackage{listings}
\usepackage{tikz,subfigure}
\usepackage{url}
\usepackage{booktabs} 
\usepackage{multirow}
\usepackage{xspace}

\usepackage{hyperref}
\usepackage{hyphenat}
\usepackage{arydshln}
\usepackage{amsthm}

\graphicspath{{Figures/}}
\input{Preamble/macros}

\DOI{foobar}

\cclogo{\includegraphics{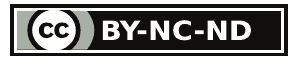}}

\begin{document}
\pagestyle{plain}

 \author*[1]{I\~{n}igo Querejeta-Azurmendi}

 \author[2]{Panagiotis Papadopoulos}

 \author[3]{Matteo Varvello}
  
 \author[4]{Antonio Nappa}

 \author[5]{Jiexin Zhang}
  
 \author[6]{Benjamin Livshits}

 \affil[1]{Universidad Carlos III Madrid / ITFI, CSIC. Part of the work performed while working at Brave Software.}

 \affil[2]{Telefónica Research}

 \affil[3]{Bell Labs}
  
 \affil[4]{University of California, Berkeley}

 \affil[5]{University of Cambridge}
  
 \affil[6]{Brave Software/Imperial College}

\title{\NAME: A Friction-less Privacy-Preserving Human Attestation Mechanism for Mobile Devices}

\keywords{Human Attestation, Privacy Preserving Bot Detection, Frictionless verification of humanness}

\begin{abstract}
{Recent studies show that~20.4\% of the internet traffic originates from automated agents. To identify and block such ill-intentioned traffic, mechanisms that \emph{verify the humanness of the user} are widely deployed, with CAPTCHAs being the most popular. 
Traditional CAPTCHAs require extra user effort (\eg solving mathematical puzzles), which can severely downgrade the end-user's experience, especially on mobile, and provide sporadic humanness verification of questionable accuracy. More recent solutions like Google's reCAPTCHA~v3, leverage user data, thus raising significant privacy concerns.
To address these issues, we present \name: the first zero-knowledge proof-based humanness attestation system for mobile devices. \name moves the human attestation to the edge: onto the user's very own device, where humanness of the user is assessed in a privacy-preserving and seamless manner. \name achieves this by classifying motion sensor outputs of the mobile device, based on a model trained by using both publicly available sensor data and data collected from a small group of volunteers. 
To ensure the integrity of the process, the classification result is enclosed in a zero-knowledge proof of humanness that can be safely shared with a remote server.
We implement \name as an Android service to demonstrate its effectiveness and practicality. In our evaluation, we show that \name successfully verifies the humanness of a user across a variety of attacking scenarios and demonstrate~\modelaccuracy accuracy. On a two years old Samsung S9, \name's attestation takes around \proverLatency (when visual CAPTCHAs need~9.8 seconds) and consumes a negligible amount of battery.}
\end{abstract}

  \journalname{Proceedings on Privacy Enhancing Technologies}
  \startpage{1}
  \journalvolume{..}

\maketitle
\vspace{-1.5cm}
\section{Introduction}
\label{sec:intro}
Automated software agents that interact with content in a human-like way, are becoming more prevalent and pernicious in the recent years. Web scraping, competitive data mining, account hijacking, spam and ad fraud are attacks that such agents launch by mimicking human actions at large scale. According to a recent study~\cite{botTraffic},~20.4\% of the~2019 internet traffic was fraudulent, associated with \emph{user, albeit not human, activity}.

In the ad market specifically, such type of fraudulent traffic costs companies between~\$6.5-\$19 billions in the U.S. alone, and it is estimated that this will grow to~\$50 billion by~2025~\cite{fraudCostForecast}. When it comes to the ever-growing mobile traffic, a recent study~\cite{mobileFraud2018} (using data spanning~17 billion transactions) observes~189 million attacks originated specifically from mobile devices; this is an increase of~12\% compared to the previous six months.

The ``Completely Automated Public Turing tests to tell Computers and Humans Apart'' (or just CAPTCHA) is the current state-of-the-art mechanism to assess the humanness of a user. CAPTCHAs are widely deployed across the internet to identify and block fraudulent non-human traffic. The major downsides of current CAPTCHA solutions (\eg Securimage~\cite{secureimage}, hCaptcha~\cite{hcaptcha}) include: {\bf (i) questionable accuracy}: various past works demonstrate how CAPTCHAs can be solved within milliseconds~\cite{captchaAccuracy,sivakorn2016robot,yan2008low,captchabreak,bock2017uncaptcha}, {\bf (ii) added friction}: additional user actions are required (\eg image, audio, math, or textual challenges) that significantly impoverish the user experience~\cite{gafni2016captcha}, especially on mobile devices, 
{\bf (iii) discrimination}: poor implementations often block access to content~\cite{captchaFP}, especially to visual-impaired users~\cite{captchaInaccess}
{\bf (iv) serious privacy implications}: to reduce friction, Google's reCAPTCHAv3~\cite{captcha_v3} replaces proof-of-work challenges with extensive user tracking. Google's servers attest user's humanness by collecting and validating behavioral data~\cite{captchaData} (\ie typing patterns, mouse clicks, stored cookies, installed plugins), thus raising significant privacy concerns~\cite{captchaPrivacy1,captchaPrivacy2,Akrout2019HackingLearning}. 


The goal of this paper is to build a humanness attestation alternative that will put an end to the false dilemma: humanness attestation at the cost of user experience or at the cost of user privacy? By leveraging the device's motion sensors we demonstrate that it is possible to build a humanness attestation mechanism on the edge that preserves the privacy of the users and runs seamlessly on the background thus requiring zero interaction from the mobile user.

The key intuition behind our approach is that whenever a (human) user interacts with a mobile device, the force applied during the touch event generates motion. This motion can be captured by the device's sensors (\eg accelerometer and gyroscope) and used to uniquely distinguish real users from automated agents. Further, this detection can run on the user device, without requiring secure execution environment (unlike related proposals~\cite{guerar2018,jamshed2010suppressing}), but more importantly without sharing private information with any server (unlike state-of-the-art~\cite{captcha_v3}), apart from a \emph{proof} that guarantees the integrity of the attestation result. Secure execution environments are rapidly evolving and might soon become a crucial tool for human attestation mechanisms, however, we have not yet seen wide adoption by low-end devices. 
We realize this vision with \name, a friction-less and privacy-preserving mechanism for humanness attestation that aims to replace CAPTCHA in mobile devices. 
This paper makes the following contributions: 

\begin{enumerate}[leftmargin=0.5cm, topsep=0pt]
    \item We design a human attestation system (\name) that is both friction-less and privacy preserving. \name leverages mobile motion sensors to verify that user actions (\eg type/touch events) on a mobile device are triggered by an actual human. Such classification takes place by studying the output of the mobile device's motion sensors during the particular user action. 
    To set the ground truth, we use publicly available sensor traces and we instrument an actual Android browser app to capture both user clicks and sensor traces from a small set of real users. 
    Our approach is tested under various scenarios: (i) when device is resting (on a table), (ii) when there is artificial movement from device's vibration,  or (iii) from an external swinging cradle.

    \item We develop  a sub-linear inner product zero-knowledge proof, which 
    we use  to build (and open-source\footnote{zkSVM source code: \url{https://github.com/zkSENSE/zkSVM}}) \innerProdName: a zero-knowledge based library for enclosing results of an SVM (Support-Vector Machine) classifier into zero-knowledge proofs. \innerProdName leverages arithmetic properties of commitment functions and prover-effective proofs to ensure the integrity of the classification result reported to a remote server.

    \item We implement an Android SDK 
    and a demo app\footnote{Demo video: \url{https://youtu.be/U-tZKrGb8L0}} to showcase the detection accuracy of \name. 
    \name is \emph{invisible to the user} and capable of verifying their humanness with accuracy higher than related proposals~\cite{guerar2018} (\modelaccuracy). Performance evaluation results of our prototype show that the entire attestation operation takes less than~\proverLatency (compared to ~9 sec~\cite{captchaUX} required by Android reCAPTCHA~\cite{android_captcha} and consumes less than~\powerOverhead of power.
    
\end{enumerate}
\section{Goals and Threat Model}
\label{sec:principles}
In this section, we describe  (i) the basic design principles that a humanness verification mechanism must follow, and (ii) how existing mechanisms work and compare with \name. Specifically, a successful human attestation mechanism must:

\begin{enumerate}[wide, labelindent=0pt, topsep=0pt, label=\textbf{\Alph*})]
\item {\bf Be friction-less:} Most existing mechanisms require the user to solve mathematical quizzes or image/audio challenges~\cite{secureimage,hcaptcha}, thus severely hampering the user experience. According to studies~\cite{bursztein2010good,captchaUX}: (i) humans only agree on what the CAPTCHA says 71\% of the time, (ii) visual CAPTCHAs take~9.8 seconds (on average) to complete, (iii) audio CAPTCHAs take 28.4 seconds (and~50\% of the Audio CAPTCHA users quit). The profound degradation of the user experience forces service providers to perform user attestations only sporadically~\cite{sporadically3,sporadically2,sporadically1} in an attempt to save their declining conversion rates. Related research works~\cite{guerar2015completely, 10.1007/978-3-319-45572-3_3, 10.1007/978-3-319-02937-5_11} reduce user friction by requiring, for example, the user to tilt their phone during humanness verification. In contrast, \name is completely friction-less: humanness is attested via device micro-movements that happen during natural user actions like  typing, and screen touching.

\item {\bf Be privacy-preserving}: To mitigate the above, mechanisms like reCAPTCHA v3~\cite{captcha_v3} (i) track the user while browsing a webpage, (ii) send raw tracking data to third party attestation servers, where (iii) a ``risk score'' representative of humanness is computed and then (iv) shared with the webmasters. Of course, this pervasive behavioral tracking raises significant privacy concerns~\cite{captchaPrivacy1,captchaPrivacy2}. Similarly, there are sensor-based approaches~\cite{guerar2018,guerar2015completely,10.1007/978-3-319-45572-3_3} 
that transmit raw mobile sensor data to remote attestation servers; something that as reported, may reveal keystrokes, gender, age, or be used to fingerprint users~\cite{das2016tracking,REYESORTIZ2016754,SANSEGUNDO2018190,Malekzadeh:2018:PSD:3195258.3195260,davarci2017age,zhang2019sensorid,2017:LPD:3038912.3052691}. Contrary to that, \name decouples the humanness attestation procedure from the server, and moves it to the edge.
By performing the entire attestation on the user's very own device, no sensitive data but the classification result (enclosed in zero-knowledge proofs to ensure integrity) leaves the user's mobile device.

\item {\bf Be broadly accessible:} The majority of CAPTCHA solutions are not accessible to all users~\cite{captchaDisadv,captchaInaccess}. According to a survey~\cite{captchaAccess}, CAPTCHAs are the main source of difficulty for visually impaired users. Meanwhile, human attestation mechanisms designed for visually impaired people (like audio-based reCAPTCHA) have been exploited to bypass CAPTCHAs by providing automatic responses~\cite{buster,tam2009breaking}. \name's seamless integration with a user's natural interaction with a device does not introduce any additional barrier to people affected by any form of disability. We must clarify however, that \name is not tested to serve people that use voice commands instead of screen touching (and thus do not cause micro-movements during scroll, clicks, \etc). Extending our classifier to work with the buttons of the devices (\ie lock/unlock, volume up/down) that visually impaired people use is part of our future work.
\end{enumerate}

\subsection{Threat Model}
\label{sec:threatModel}
\point{Formal security properties} In Appendix~\ref{sec:formal-analysis} we introduce formal notions of privacy and verifiability, and prove that \innerProdName provides them. Informally, these notions ensure that the user data is kept private from an adversary, and that a proof convinces a verifier that the submitted result can only be originated by applying the SVM model to the \emph{committed} vector. \name does not provide any guarantees on the origin of the input data itself. Hence, as with other widely deployed human attestation mechanisms (\ie reCAPTCHAv3) an adversary capable of forging human activity is capable of bypassing the mechanism. Our scheme provides replay prevention of proofs, but not of the data used to generate such proofs, making it possible to perform several attestations with a single vector. 

\point{Attacker}  
We assume the same attacker model as in CAPTCHA systems: an attacker who is capable of automating user actions while accessing an online service. 
The goal of such attacker is to imitate a legitimate user and launch attacks like content scraping, ad viewing/clicking, API abuse, spam sending, DDoS performing, \etc, for monetary gain. This monetary gain can be achieved either (i) indirectly: \eg app developers pay attackers to perform Black Hat App Store Optimization attacks to increase their app ranking~\cite{shuabang1,blackaso,shuabang2},
or (ii) directly: an attacker registers for reward schemes in mobile apps (\eg Brave's Ad Rewards~\cite{braveRewards}) via multiple accounts and claim rewards. 

As for CAPTCHA systems, click farms are not included in this threat model, since they rely on malicious but human activity. Additionally, we assume that the attacker does not control the device's OS (rooted device). Such a powerful attacker has the capabilities to launch much more severe attacks (\eg install spyware, steal passwords or sensitive user data, modify firmware, tamper with sensors)~\cite{stealpwd,pegasus}. 

\point{Server-Auditor} 
We assume a set of servers that act as \emph{auditors}, simply receiving and verifying the humanness classification results the users report. Contrary to reCAPTCHAv3, such servers are not required to be trusted by the users. Users can report a server that denies issuing tokens to an attested user.

\point{User} 
Throughout the rest of this paper, when we mention user activity, we mean screen interaction, which results (when interpreted by the mobile OS) in clicking, key typing, or scrolling. We further assume mobile devices equipped with a set of sensors that includes a gyroscope and an accelerometer. 
\section{Building Blocks}
\label{sec:background}

In this section, we briefly introduce the concepts of zero-knowledge proofs and commitment functions: the basic pillars of our privacy preserving construction.

\subsection{Preliminaries}
Let $\group$ be a cyclic group with prime order $\order$ generated by generator $\generator$. Let $\hgenerator$ be another generator of the group $\group$ where the discrete logarithm of $\hgenerator$ with base $\generator$ is not known. Furthermore, let $\generator_1, \ldots, \generator_{\lastgenerator}, \hgenerator_1, \ldots, \hgenerator_{\lastgenerator}$ denote $2\lastgenerator$ distinct group elements where their relative discrete logarithms are unknown. We denote the integers modulo $\order$ as $\Zp$ and write $a\randin S$ to denote that $a$ is chosen uniformly at random from a set $S$.  Although any prime order group where the Decisional Diffie–Hellman~(DDH) assumption holds may be used, in our implementation we used the ristretto255 group~\cite{ristretto255} over Curve25519~\cite{curve25519-dalek}.

\point{Notation} We use bold letters to denote vectors. In particular, $\vectorgenerator \in \group^{\lastgenerator}$ is defined by $(\generator_1, \ldots, \generator_{\lastgenerator})$ with $\generator_i\in\group$, and $\bm{a}\in\Zp^{\lastgenerator}$ by $(a_1, \ldots, a_{\lastgenerator})$ with $a_i\in\Zp$. We use normal exponentiation notation to denote the multi-exponentiation of two vectors, with $\vectorgenerator^{\bm{a}} = \prod_{i=1}^{\lastgenerator}\generator_i^{a_i}$. Moreover, two vector multiplications (or additions) represents the entry wise multiplication (or addition). 




\subsection{Zero-Knowledge Proofs}
\label{sec:zkp-background}
In~\cite{Goldwasser:1985:KCI:22145.22178} the technique of (interactive) zero-knowledge Proof (ZKP) is presented to enable one party (prover) to convince another (verifier) about the validity of a certain statement. The proof must provide the following informal properties (we formalise them in Appendix~\ref{appendix:proof}): 
\begin{itemize}[topsep=0.1cm]
    \item Completeness: If the prover and verifier follow the protocol, the latter always validates.
    \item Knowledge soundness: The verifier only accepts the proof if the statement being proven holds.
    \item Honest-verifier zero-knowledge: The prove discloses no other information other than the fact that the statement is true. 
\end{itemize}
Blum~\etal~\cite{Blum:1988:NZA:62212.62222} introduced non-interactive zero-knowledge Proofs~(NIZKPs), which enable the prover to prove the validity of a statement without interacting with the verifier. NIZKPs enjoy a growing popularity and adoption in the blockchain era, used across various applications mostly for decentralization, verifiability and accountability~\cite{varvello2019vpn0,zerocash, icash, pestana2020themis}. NIZKPs can be extended to Signature Proofs of Knowledge (SPK)~\cite{Camenisch:1997:EGS:646762.706305}, in which the prover signs a message while proving the statement. 
We adopt the Camenisch-Stadler notation~\cite{Camenisch:1997:EGS:646762.706305}
to denote such proofs and write, for example:\vspace{-0.2cm}
\newcommand{\challengeproof}{w}
\begin{equation*}
  \SPK\{ (x) : A = \generator^{x} \;\land\; B = A^x \}(\challengeproof)
\end{equation*}
to denote the non-interactive signature proof of knowledge, over message $\challengeproof$, that the prover knows the discrete log of $A$ and $B$ with bases $\generator$ and $A$ respectively, and that the discrete log is equal in both cases. Values in the parenthesis are private (called the witness), while all other values in the proof are public. We omit the notation of the signed message throughout the whole paper, but we assume that all \SPK s sign a challenge sent by the server every time it requests human attestation. This prevents replay attacks of zero knowledge proofs. We represent proofs with $\Pi$, and define two functions: the generation function, $\proofgen{\Pi}()$, and the verification function, $\proofverif{\Pi}()$. The inputs are, for the case of the generation, the public values and the witness (in the example above, $A, B$ and $x$). On the other hand, the verification function only takes the public values as input (in the example above, $A$ and $B$). 

Blockchain has created a high interest in building proofs with minimal size and verifier computation~\cite{Groth16, Sonic, Plonk, bulletproofs, BCC+16}, but with the cost of increasing the prover's time. In Section~\ref{sec:evaluation}, we show how proving execution of the SVM model based on~\cite{Groth16} (one of the most widely used such proofs) is unbearable for mobile devices. 

\subsection{Homomorphic Commitment Functions}
\label{sec:crypto-background}
\newcommand{\mess}{m}
\newcommand{\vectormess}{\bm{\mess}}
\newcommand{\blindingfactor}{r}
\newcommand{\opening}{\texttt{Open}}
In this paper, we use additive homomorphic commitment functions, a cryptographic primitive that allows a party to commit to a given value, providing the hiding property (the value itself is hidden) and the binding property (a commitment cannot be opened with a different value than the original one). We present our work
based on the Pedersen Commitment Scheme~\cite{pedersencomm}. The Pedersen Commitment scheme takes as input message $\mess\in\Zp$ and a hiding value $\blindingfactor\randin\Zp$ and commits to it as follows:
\[\commit(\mess, \blindingfactor) = \generator^{\mess}\cdot \hgenerator^{\blindingfactor} .\]
Note the additive homomorphic property, where:
\begin{multline*}
\commit(\mess_1, \blindingfactor_1)\cdot\commit(\mess_2, \blindingfactor_2) =  \\ \generator^{\mess_1 + \mess_2}\cdot \hgenerator^{\blindingfactor_1 + \blindingfactor_2} = \commit(\mess_1 + \mess_2, \blindingfactor_1 + \blindingfactor_2).
\end{multline*}
Furthermore, in this paper we use a generalization of the Pedersen Commitment \cite{veccommitments}, that instead takes a message in $\vectormess\in\Zp^{\lastgenerator}$ and a hiding value $\blindingfactor\randin\Zp$ and commits to it as follows:
\[\commit(\vectormess, \blindingfactor) = \vectorgenerator^{\vectormess}\hgenerator^{\blindingfactor} =  \prod_{i=1}^{\lastgenerator}\generator_i^{\mess_i}\cdot\hgenerator^{\blindingfactor}.\]
\noindent Note that such a commitment function has the same additive property as above, where addition in the message committed happens entry wise. Finally, a commitment function has an opening verification function, where, given the opening and blinding factor, determines whether these correspond to a given commitment,~$C$. More specifically:
\[
\opening(C, \vectormess, \blindingfactor, \vectorgenerator, \hgenerator) = 
\begin{cases}
\top & \text{if } C=\vectorgenerator^{\vectormess}\cdot\hgenerator^{\blindingfactor} \\
\perp & \text{else}
\end{cases}
\]

We leverage zero-knowledge proofs over Pedersen commitments. First, we use a proof of knowledge of the opening of a commitment~\cite{veccommitments}, $\proofopening$, and a proof of equality, $\proofequality$. Particularly:
\begin{equation*}
\proofopening := \SPK\left\{ (\vectormess, \blindingfactor): 
\commitment =  \vectorgenerator^{\vectormess}\cdot\hgenerator^{\blindingfactor}\right\}.
\end{equation*} 
and 
\begin{multline*}
\proofequality := \SPK\left\{(\vectormess, \blindingfactor_1, \blindingfactor_2): \right. \\ \left.
\commitment_1 =  \vectorgenerator^{\vectormess}\cdot\hgenerator^{\blindingfactor_1} 
\wedge 
\commitment_2 =  \vectorhgenerator^{\vectormess}\cdot\hgenerator^{\blindingfactor_2}\right\},
\end{multline*} 
respectively.
We also construct a ZKP that allows us to provably exchange any element of the committed vector by a zero. To this end we provably extract the particular value we want to replace with its corresponding  generator, and divide it from the initial commitment. More particularly:
\begin{multline*}
\proofzero := \SPK\left\{(\vectormess\in\Zp^{\lastgenerator}, \blindingfactor): 
\commitment =  \vectorgenerator^{\vectormess} \hgenerator^{\blindingfactor} \wedge \extracted = \generator_j^{\mess_j} \wedge \right.\\
\left.\commitment / \extracted = \generator_1^{\mess_1} \cdots \generator_{j-1}^{\mess_{j-1}}\cdot
\generator_{j+1}^{\mess_{j+1}} \cdots \generator_{\lastgenerator}^{\mess_{\lastgenerator}} \hgenerator^{\blindingfactor}\right\}.
\end{multline*}

\noindent Next, we use the proof that a committed value is the square of another committed value presented in~\cite{proofsquare}:
\begin{multline*}
\proofsquare := \SPK\left\{(\mess_1, \mess_2, \blindingfactor_1, \blindingfactor_2 \in\Zp):\right. \\
\left.\commitment_1 = \generator^{\mess_1}\cdot\hgenerator^{\blindingfactor_1} \wedge \commitment_2 = \generator^{\mess_2}\cdot\hgenerator^{\blindingfactor_2} \wedge \mess_1 = \mess_2 ^ 2\right\}.
\end{multline*} 

\noindent To prove that a number is within a range, we leverage Bulletproofs~\cite{bulletproofs}:\vspace{-0.1cm}
\begin{equation*}
\proofge := \SPK\left\{(\mess,\blindingfactor\in\Zp):\commitment = \generator^{\mess}\hgenerator^{\blindingfactor} \wedge \mess \in [0, 2^l]\right\}
\end{equation*}
for some positive integer $l$. We combine these two proofs, $\proofsquare$ and $\proofge$, to prove the correct computation of the floor of the square root of a committed value. Let $\commitment_1 = \generator^{\mess_1}\cdot\hgenerator^{\blindingfactor_1}$, $\commitment_2 = \generator^{\mess_2}\cdot\hgenerator^{\blindingfactor_2}$ and $\commitment_1^{+1}=\commitment_1\cdot\generator = \generator^{\mess_1 + 1}\cdot\hgenerator^{\blindingfactor_1}$. We want to prove that $\mess_1 = \lfloor\sqrt{\mess_2}\rfloor$. It suffices to prove the following two statements: 
\begin{itemize}[topsep=0pt]
    \item[(i)] The square of the committed value in $\commitment_1$ is smaller or equal than $\commitment_2$, and  
    \item[(ii)] The square of the committed value in $\commitment_1^{+1}$ is greater than $\commitment_2$.
\end{itemize}
We denote this proof by:\vspace{-0.3cm}
\begin{multline*}
    \proofsqrt := \SPK\left\{(\mess_1, \mess_2, \blindingfactor_1, \blindingfactor_2 \in\Zp):\right. \\
\left.\commitment_1 = \generator^{\mess_1}\cdot\hgenerator^{\blindingfactor_1} \wedge \commitment_2 = \generator^{\mess_2}\cdot\hgenerator^{\blindingfactor_2} \wedge \mess_1 = \lfloor\sqrt{\mess_2}\rfloor\right\}.
\end{multline*}

\section{Sub-linear Inner Product Zero-Knowledge Proof}
\label{sec:zk-ip}
\newcommand{\ipzkp}{IP-ZKP\xspace}

\renewcommand{\proof}{\pi}

\newcommand{\commitmentvectors}{A}
\newcommand{\commitmentinnerproduct}{V}
In this section, we present the Sub-linear Inner Product Zero-Knowledge Proof (IP-ZKP). Our construction is based on the non zero-knowledge version presented in \cite{bulletproofs}. For their use cases, the zero-knowledge property is not required, and hence the lack of an explicit definition and study of a zero-knowledge proof. In this section we make it explicit, and in Appendix~\ref{appendix:proof} we prove that it provides completeness, knowledge soundness and special honest-verifier zero-knowledge properties. In our construction the prover has a commitment, $\commitmentvectors=\hgenerator^{\alpha}\vectorgenerator^{\leftvector}\vectorhgenerator^{\rightvector}$, of two vectors $\leftvector, \rightvector$ with blinding factor $\alpha$, and a second commitment $\commitmentinnerproduct=\generator^{\innerproduct}\hgenerator^{\gamma}$, of a value $\innerproduct$, with blinding factor $\gamma$. The prover convinces the verifier that $\langle \leftvector, \rightvector \rangle = \innerproduct$ holds. 
The prover and verifier interact as follows (can be made non-interactive with the Fiat-Shamir Heuristic~\cite{10.1007/3-540-47721-7_12}): 

\begin{figure*}[t]
    \centering
    \vspace{-.2cm}
  \includegraphics[width=0.9\linewidth]{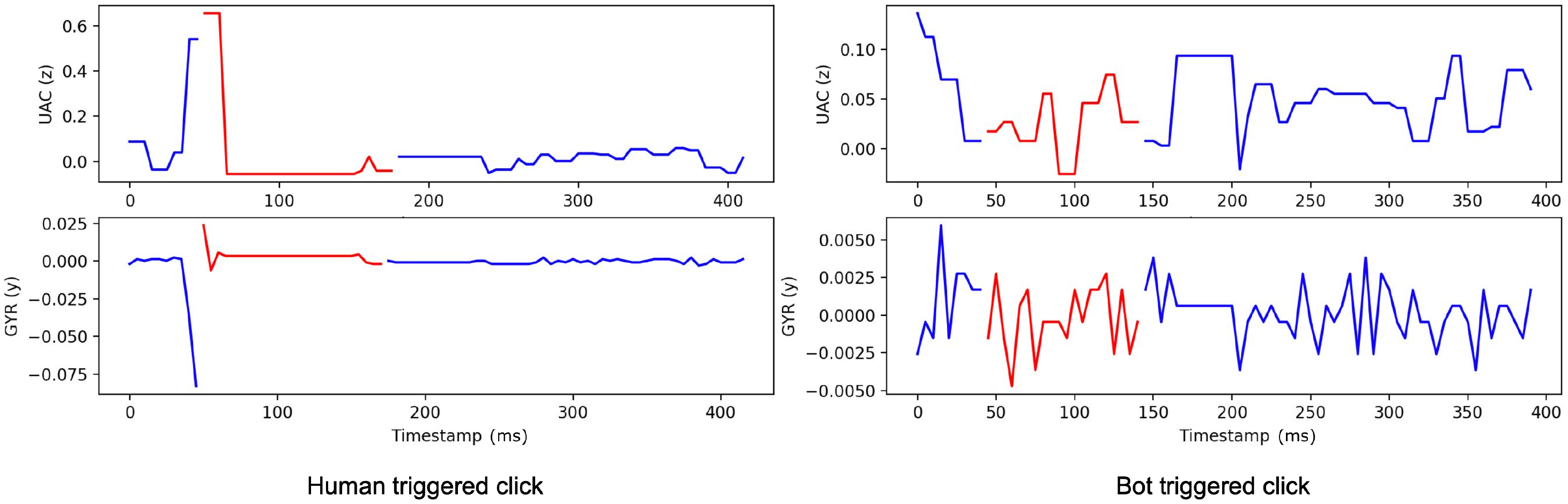}
  \vspace{-.2cm}
  \caption{Output of gyroscope and accelerometer motion sensors during human and automatically triggered click (red). The maximum linear acceleration movement is up to~8.5$\times$ greater and the angular rotational velocity is up to~4.9$\times$ greater in case of a human triggered click.}
  \label{fig:sensorsOutput}
\end{figure*}
\vspace{-.2cm}
\begin{figure*}[t]
    \centering
    \vspace{-.2cm}
  \includegraphics[width=0.9\linewidth]{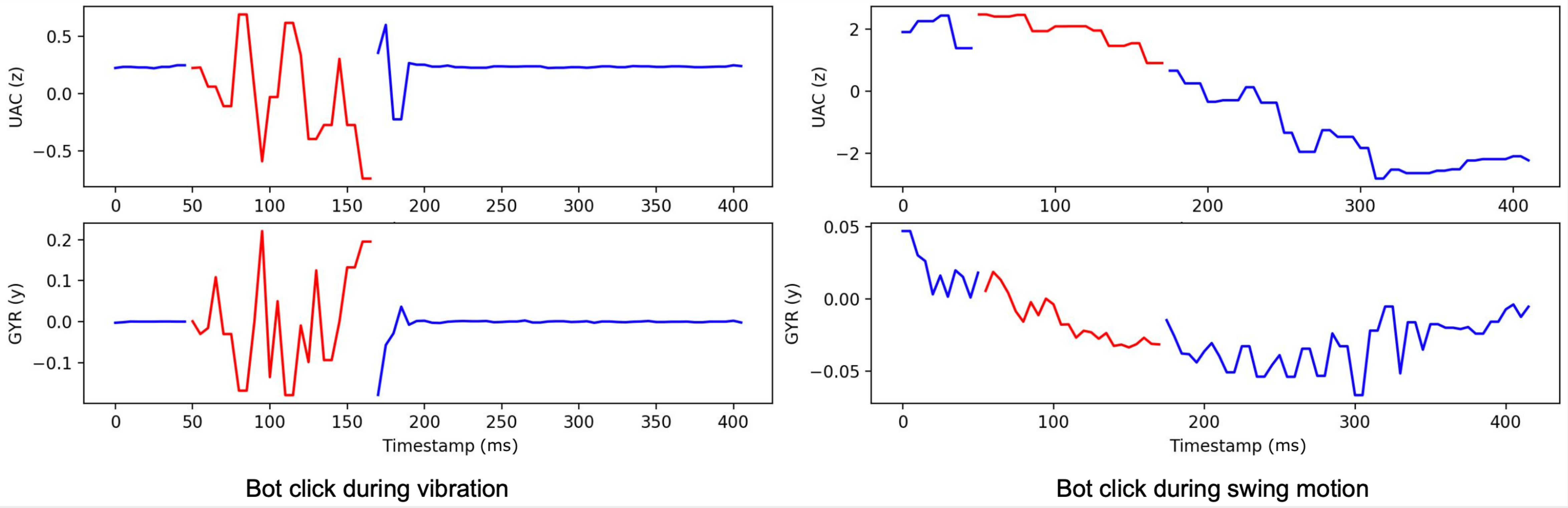}
  \vspace{-.2cm}
  \caption{Motion sensors output during automatically triggered click with artificial device movement (red): (i) during device vibration (on the left) and (ii) when device is docked on a swing (on the right).}
  \label{fig:sensorsOutput_motion}
\end{figure*}
\begin{description}
\item[$\prover$:] It computes blinding vectors, $\blindingvectorleft, \blindingvectorright$ for each vector of the inner product and commits to it:
\vspace{-0.3cm}
\begin{align}
\blindingvectorleft, \blindingvectorright &\randin\Zp^{\lastgenerator}\\
\blindingblindings &\randin\Zp \\
\commitmentblinders &= \hgenerator^{\blindingblindings}\vectorgenerator^{\blindingvectorleft}\vectorhgenerator^{\blindingvectorright}\in\group
\end{align}
\sloppy Now the prover defines two linear vector polynomials, $\leftpolynomial(\variable), \rightpolynomial(\variable)\in\Zp^{\lastgenerator}[\variable]$, and a quadratic polynomial as follows: 
\begin{align}
\hspace{2mm}\leftpolynomial(\variable) &= \leftvector + \blindingvectorleft\cdot\variable \in\Zp^{\lastgenerator}[\variable] \\
\rightpolynomial(\variable) &= \rightvector + \blindingvectorright\cdot\variable \in\Zp^{\lastgenerator}[\variable] \\
\begin{split}
\innerproductpolynomial(\variable) &= \langle\leftpolynomial(\variable), \rightpolynomial(\variable)\rangle = \\
& \hspace{1cm}\coefficientip{0} + \coefficientip{1}\cdot\variable + \coefficientip{2}\cdot\variable^2\in\Zp^{\lastgenerator}[\variable]
\end{split}
\end{align}
By creating like that the polynomials, it allows for an evaluation of the polynomial at a given point without disclosing any information about the vectors $\leftvector$ or $\rightvector$. The prover needs to convince the verifier that the constant term of $\innerproductpolynomial(\variable)$ equals $\innerproduct$. To do so, the prover commits to the remaining coefficients of the polynomial
\begin{align}
\blindingcoefficient_1, \blindingcoefficient_2 &\randin\Zp \\
\commitmentcoefficients_i &= \generator^{\coefficientip{i}}\hgenerator^{\blindingcoefficient_i}\in\group, i=\lbrace 1, 2\rbrace
\end{align}
\item[$\prover\rightarrow\verifier$:] $\commitmentblinders, \commitmentcoefficients_1, \commitmentcoefficients_2$
\item[$\verifier$:] $\challenge\randin\Zp^*$
\item[$\verifier\rightarrow\prover$:]$\challenge$
\item[$\prover$:] It computes the response using the challenge received
\begin{align}
\evaluatedleftpolynomial &= \leftpolynomial(\challenge) = \leftvector + \blindingvectorleft \cdot \challenge \in\Zp^{\lastgenerator} \\
\evaluatedrightpolynomial &= \rightpolynomial(\challenge) = \rightvector + \blindingvectorright\cdot \challenge \in \Zp^{\lastgenerator} \\
\evaluatedinnerproductpolynomial &= \langle \evaluatedleftpolynomial, \evaluatedrightpolynomial\rangle\in\Zp \\
\blindingevalippoly &= \blindingcoefficient_2\cdot\challenge^2 + \blindingcoefficient_1\cdot\challenge + \gamma \\
\blindingannouncement &= \alpha + \blindingblindings \cdot \challenge \in\Zp
\end{align}
\item[$\prover\rightarrow\verifier$:] $\blindingevalippoly, \blindingannouncement, \evaluatedinnerproductpolynomial, \evaluatedrightpolynomial, \evaluatedleftpolynomial$. 
\item[$\verifier$:] The verifier needs to check that $\evaluatedrightpolynomial, \evaluatedleftpolynomial$ are correct and the inner product relation holds with respect to $\evaluatedinnerproductpolynomial$. To this end it performs the following checks: 
\begin{align}
\generator^{\evaluatedinnerproductpolynomial}\hgenerator^{\blindingevalippoly} &\stackrel{?}{=}\commitmentinnerproduct\cdot\commitmentcoefficients_1^{\challenge}\cdot\commitmentcoefficients_2^{\challenge^2} \label{eq:commitments-polys}\\
P &= \commitmentvectors\cdot\commitmentblinders^{\challenge}\in\group\\
P &\stackrel{?}{=} \hgenerator^{\blindingannouncement}\cdot\vectorgenerator^{\evaluatedleftpolynomial}\cdot\vectorhgenerator^{\evaluatedrightpolynomial} \label{eq:p-equality}\\
\evaluatedinnerproductpolynomial &\stackrel{?}{=}\langle\evaluatedleftpolynomial, \evaluatedrightpolynomial \rangle
\end{align}
If all checks validate, then this means that the statement is true with very high probability. 
\end{description}
To make this proof logarithmic, we use the same trick as in the original paper. Instead of sending the vectors $\evaluatedrightpolynomial, \evaluatedleftpolynomial$ to prove the inner product relation, we leverage IP-ZKP over the blinded vectors recursively. 
We prove the following Theorem in Appendix~\ref{appendix:proof}:
\begin{theorem}
\label{thm:ip-zkp}
The inner product proof presented in Protocol 1 has perfect completeness,
perfect special honest verifier zero-knowledge, and knowledge soundness.
\end{theorem}\vspace{-0.2cm}

\begin{figure*}[t]
    \centering
    \includegraphics[width=0.8\textwidth]{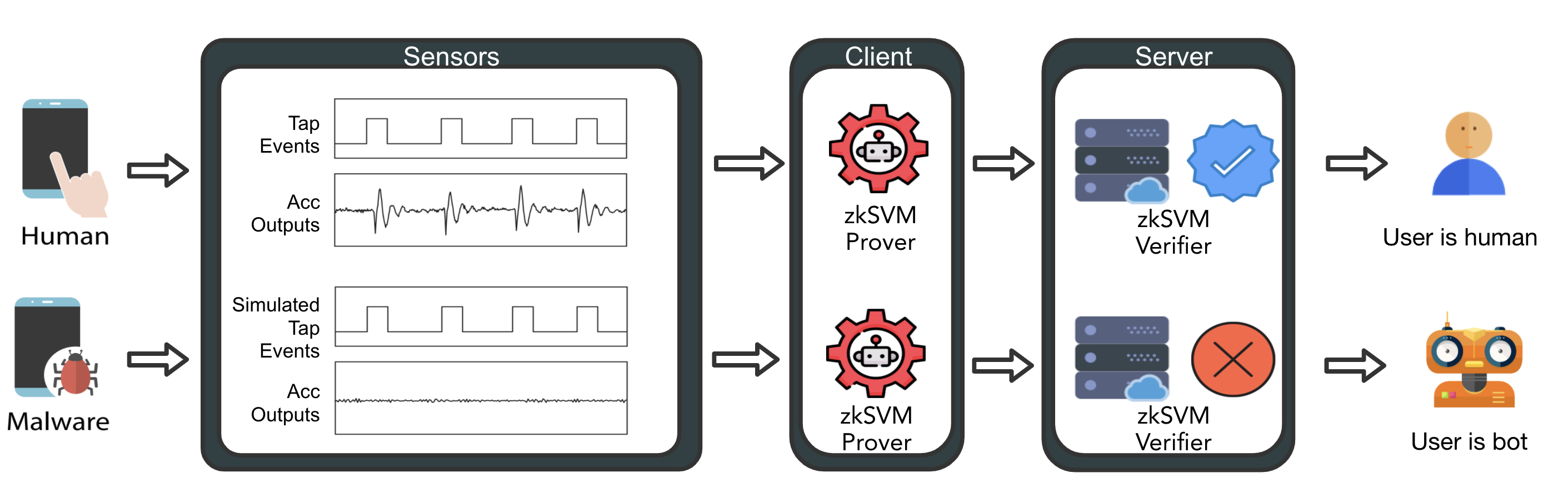}\vspace{-0.4cm}
    \caption{High-level overview of the \name architecture. An integrated ML-based classifier studies the patterns of sensor outputs right before, during, and shortly after a click event. To avoid leaking sensitive sensor output outside the device, the classification appears on the user side and the client has to prove the integrity of the reported result to the server.}\vspace{-0.2cm}
    \label{fig:overview}
\end{figure*} 

\section{\NAME Overview}
\label{sec:design}
The key intuition behind \name is that whenever a (human) user 
interacts with the mobile's display, the force applied during the touch event generates motion. This motion is captured by the embedded IMU (Inertial Measurement Unit) sensors (\eg accelerometer and gyroscope). By contrast, when there is automated user activity (\eg simulated touches) there is no external force exerted by fingers, and thus there is no noticeable change in the output of the above sensors. 

Figure~\ref{fig:sensorsOutput} shows a snapshot of the output produced by IMU sensors during click events performed both by a human (left plots) and an automated agent (right plots). 
During the click events (highlighted in red) the  accelerometer (top plots) senses a max rate of change of~0.6 in case of human and~0.07 for an automated agent, \ie ~8.5$\times$ greater maximum linear acceleration movement. Similarly, the gyroscope (bottom plots) senses a max rate of change of~0.024 in case of human click and~0.0049 when there is automation, \ie~4.9$\times$ greater maximum angular rotational velocity.

Figure~\ref{fig:sensorsOutput_motion} shows a snapshot of the same sensors' output in the case of automated clicks coupled with two artificial device movements: vibration and swing. With vibration (left plots), we see that the motion generated is comparable with the case of the human click depicted in Figure~\ref{fig:sensorsOutput}. 
We see that the accelerometer senses the same force with the case of the human click but for a longer time -- this verifies the observations of~\cite{guerar2018}. The gyroscope though senses greater angular rotational velocity and for longer time than in the case of a human's click. 
With swing (right plots), the gyroscope senses similar angular rotational velocity as with the case of the human click, while the accelerometer senses greater linear acceleration movement (up to~3.8$\times$) for a long period.

\subsection{System Overview}
\label{sec:overview}
Building upon the above observations, \name uses an ML-based classifier to study the pattern of sensor outputs before, during, and shortly after a click event (see Figure~\ref{fig:overview}). Based on this information, the model decides about whether the action was triggered by a human or not.


\point{Human Attestation on the Edge} Privacy-preserving evaluation of ML models has become increasingly important especially in the post GDPR era. One approach 
consists in encrypting the data on the client, and runs ML model on such encrypted data at the server. 
This can be achieved via Fully Homomorphic Encryption (FHE)~\cite{Dowlin:2016:CAN:3045390.3045413,mlconfidential,bos2013private}: clients encrypt their data with their own keys and send the ciphertext to the server to evaluate an ML model. 
Next, the server sends the outcome of the homomorphic computation back to the user who would provably decrypt it and send back to the server. It is easy to anticipate that the overhead to perform such multi-step operation for each client is unbearable for services with millions of clients. 

In \name, we pre-train a model on a server and we move the classifier to the edge by running it on the user side and only report the result to an attestation server responsible for auditing the humanness of the user. 
This way, \name ensures that private sensor data never leaves the user's device. In Figure~\ref{fig:overview}, we present the high level overview of our approach. As we can see, an attestation starts with a click (screen touch). The motion sensor outputs generated during this event are used as input to the zkSVM Prover module, which runs a trained model to classify if the action was conducted by a human or not. 

\begin{table}[t]
    \centering
    {\scriptsize
    \begin{tabular}{ll}
    \toprule
        {\bf Data} & {\bf Data amount} \\ \midrule
        Volunteering Users &  \realusers users\\
        Duration of collection & \daysCollected days \\
        Android Devices tested & Google Pixel 3,  Realme X2 Pro,\\
        & Samsung Galaxy S9/S8/S6, Honor 9, \\ 
        & Huawei Mate 20 Lite, OnePlus 6   \\
        Human events collected & \humanclicks clicks \\
        Artificial events collected & \botclicks clicks\\ \bottomrule
    \end{tabular}
    }\vspace{-0.1cm}
    \caption{Summary of the collected dataset.}\vspace{-0.4cm}
    \label{tab:dataset}
\end{table}

\subsection{Classification of Humanness} 
\label{sec:classification}
\point{Data Collection}
To collect the necessary ground truth to train the various tested models, we instrumented the open source browser Brave for Android to capture click events (and their corresponding motion sensor traces) performed during browsing\footnote{We chose a real app to make sure that the volunteers will continue using it for a longer period contrary to a possible instrumented toy app.}. Then, we recruited \realusers volunteers\footnote{For a production ready system a larger pool of real users will need to contribute data, but for our study we found it  difficult to recruit users who would contribute their clicks for a long time.} who used our instrumented browser for \daysCollected consecutive days for their daily browsing\footnote{During data collection the instrumented browser was running on the users'  devices so we could not control their background running services. The data collected included only the raw sensor data during a user click.}. The device models used are: Google Pixel 3, Samsung Galaxy S9, S8 and S6, OnePlus 6, Realme X2 Pro, Huawei Mate 20 Lite and Honor 9. Volunteers were well-informed about the purpose of this study and gave us consent to collect and analyse the motion sensor traces generated during their screen touch events. We urged volunteers to use their phone as normal. 

To generate artificial user traffic, we used \texttt{adb}~\cite{adb} to automate software clicks on 4 of the volunteering devices. To test different attack scenarios, during the automation, we generate software clicks with the device being in 4 different states:
\begin{enumerate}[leftmargin=0.5cm, topsep=0pt]
    \item while resting on a platform (desk/stand)
    \item while being carried around in pocket
    \item while being placed on a swing motion device
    \item while device is vibrating (triggered by \texttt{adb})
\end{enumerate}
As summarized in Table~\ref{tab:dataset}, by the end of the data collection, we had~\humanclicks human-generated clicks and~\botclicks artificially generated clicks.

\begin{table}[t]
    \center
    {\footnotesize
    \setlength{\tabcolsep}{2pt}
   \begin{tabular}{lrr}
   \toprule
   {\bf Classifier} & {\bf F1 (weighted)} & {\bf Recall} \\ \midrule
        SVM & 0.92 & 0.95 \\
        Decision Tree (9 Layers) & 0.93 & 0.95 \\
        Random Forest (8 Trees, 10 Layers) & 0.93 & 0.95 \\
        KNN & 0.92 & 0.93 \\
        Neural Network (Linear Kernel) & 0.86 & 0.95 \\
        Neural Network (ReLU Kernel) & 0.91 & 0.96 \\
        \bottomrule
   \end{tabular}
   }
   \caption{Accuracy of the various tested classifiers.}\vspace{-0.5cm}
   \label{tbl:classifiers_acc}
\end{table}

\begin{figure}[t]
    \centering\vspace{-0.2cm}
    \includegraphics[width=.48\textwidth]{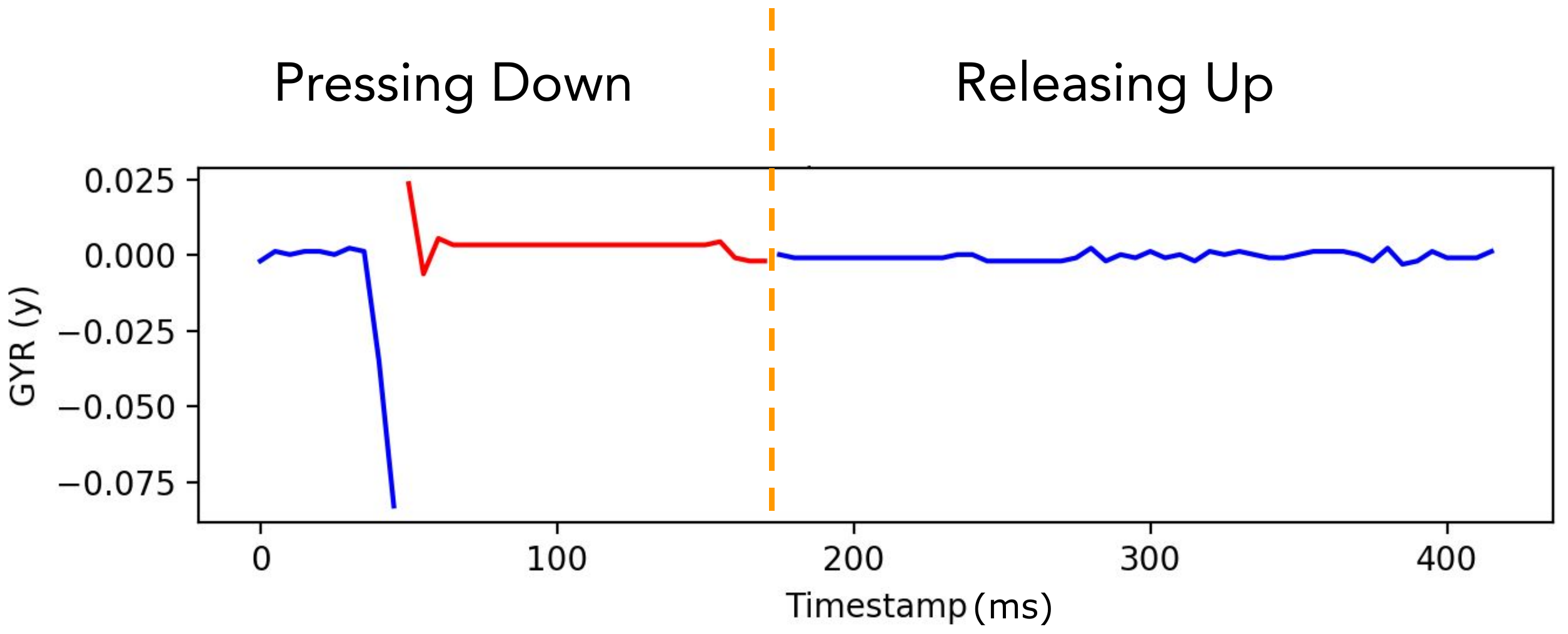}\vspace{-0.2cm}
    \caption{The period of a click event starts 50ms before the beginning of the action and ends 250ms after the end of it.}\vspace{-0.2cm}
    \label{fig:segments}
\end{figure} 

\point{Feature Selection}
\label{sec:feature-selection}
During  data collection, accelerometer and gyroscope sensors were sampled at 250Hz. For each click event, we not only consider the device motion during the touch, but also the device motion right before and shortly after the touch. In particular, we consider that the period starts 50ms before the finger touches the screen and finishes 250ms after it. Then, we split each period into two segments: (i) \emph{before releasing finger} and (ii) \emph{after releasing finger} as depicted in Figure~\ref{fig:segments}. For each axis (x, y, z) in accelerometer and gyroscope, we calculate the average and standard deviation of its outputs in each segment. In addition, we calculate the consecutive difference of sensor outputs in each segment and use the average and standard deviation of these differences as features.

\point{Classification Accuracy} 
Using the above features, we test several ML classifiers via 10-fold cross validation. Table~\ref{tbl:classifiers_acc} presents the weighted $F_1$ score and recall of the different classifiers we tested. We choose weighted $F_1$ score as an evaluation index because our dataset is unbalanced. In this context, \emph{recall} means the proportion of correctly identified artificial clicks over all artificial clicks. In other words, recall indicates the ability to capture artificial clicks. 

As shown in Table~\ref{tbl:classifiers_acc}, the four tested classifiers (\ie SVM, decision tree, random forest, and neural network with ReLU kernel) have similar performance in terms of recall (\ie 0.95 recall). Although, Decision Tree and Random Forest perform slightly better in terms of accuracy, \name utilizes SVM for compatibility purposes as stated in more detail in Section~\ref{sec:provableML}. Hence, our \name's accuracy in assessing the humanness of a user is \modelaccuracy.


\subsection{Privacy-Preserving and Provable ML}
\label{sec:provableML}
To preserve privacy, human attestation in \name is performed in the user's device and only the \emph{result} of the attestation is shared with the server. 
To ensure that the server can verify the integrity of the transmitted result, the user includes a commitment of the sensors, together with a proof that the result corresponds to the model evaluated over the committed values.

We build the two ML evaluation proving components of \name: (i) \innerProdName Prover and (ii) \innerProdName Verifier. The \innerProdName Prover checks on the client whether a user is a human based on a model we pre-trained (Section~\ref{sec:classification}), and generates a proof to ensure its proper execution.
%
%
The \innerProdName Verifier, on the server's side,
checks that the proof is correctly generated. If the verification is successful, the server will know that (a) the ML-based humanness detection model classifies the user as human or non-human based on the committed sensor outputs, and that (b) the used model is the genuine one, without though learning the value of sensor outputs.


\point{SVM enclosed in Zero-Knowledge Proofs}
\label{sec:ml_model}
Section~\ref{sec:classification} shows that the different classifiers tested achieve similar accuracy. While decision trees, random forests, or neural networks provide slightly higher F1 accuracy than SVM (see Table~\ref{tbl:classifiers_acc}), in \name we choose SVM as the underlying model due to its simplicity at evaluation time and its suitability with zero-knowledge proofs.
Contrarily, neural networks need to perform non-linear operations, while decision trees require several range proofs, which are expensive operations to prove in zero-knowledge. 
As mentioned in Section~\ref{sec:classification}, the SVM model we trained uses as features the average ($\mu$) and standard deviation ($\sigma$) of sensor outputs, together with the average and standard deviation of the consecutive difference vector. On top of that, before applying the SVM model, the extracted features need to be normalised. The goal of normalisation is to change data values to a common scale, without distorting differences in the ranges of values. Then, trained SVM weights are assigned to each normalised feature to calculate the SVM score.

Suppose for each feature $f_i$, the normalisation mean, normalisation scale, and SVM weight and intercept are $M_i$, $S_i$, $w_i$ and $c$ respectively. Then, the SVM score $s$ can be calculated with the equation:
\begin{equation}
\label{eq:svm-score}
    s = \frac{1}{e^{-(c + \sum_{i=1}^{N} \frac{(f_i - M_i)}{S_i} w_i)} + 1}.
\end{equation}
Since only the value of $f_i$ is secret, we only need to prove the computation of $\sum_{i=1}^{N} f_i \frac{w_i}{S_i}$. Given only integer values can be processed in the underlying group arithmetic of Pedersen commitments, we instead prove $\sum_{i=1}^{N} f_i \floor*{\frac{w_i}{S_i} 10^d}$ and effectively use the parameter $d$ to preserve $d$-digits after the decimal points of $\frac{w_i}{S_i}$.

\point{Building \innerProdName}
Without loss of generality, we assume that the number of sensor inputs is $\sizevector$ for every sensor. 
The protocol is divided in three phases. First, the setup phase, $\setup(\secparam)$, where the server generates the model and their corresponding weights. Secondly, the proving phase, where the prover fetches the SVM related data, computes the difference vector, and proceeds to provably compute the average and standard deviation of these values. It then applies the corresponding linear combinations to the hidden features, and opens the result to send it to the verifier. Finally, the verification phase, where the verifier checks that all computations were correctly performed. Then, it checks whether the scores do correspond to a human or a bot. 

\point{Setup phase} \noindent The setup involves the server, where it generates the cryptographic material, and trains the SVM model. The details on how this model is generated falls out of the scope of \innerProdName, the only requirement is that it follows the specifications described at the beggining of this section. Once these values are computed, they are published to allow provers to fetch them and generate the proofs.

\theoremstyle{definition}
\newtheorem{proc}{Procedure}
\begin{proc}(Setup)
\label{proc:setup}
On input the security parameter, $\secparam$, the \innerProdName server runs $\setup(\secparam)$ to generate the SVM and cryptographic parameters. Precisely, it trains the SVM model and generates the normalisation mean, normalisation scale, and SVM weight and intercept, $M_i$, $S_i$, $w_i$ and $c$ respectively. It also defines the size of the input vectors, $\sizevector$, which defines how long a touch is considered and the measurement frequency. Next it selects the group, $\group$ with generators $\generator, \hgenerator$ and prime order $\order$. It proceeds by computing two vectors, $\vectorgenerator, \vectorhgenerator$, of generators that act as bases for the Pedersen Vector Commitments. Note that the corresponding discrete log of these bases must remain unknown.
\end{proc}


\vspace{-0.3cm}
\point{Proof generation} \noindent The proof generation is divided in five protocols. First the prover computes the difference vectors and proves correctness. Next it computes the average of all vectors, followed by a computation of the standard deviation. Finally, it evaluates the normalising linear computations over the results, and sends the opening of the result to the verifier together with the proofs of correctness. The prover's secret is $\sensorvector\in\Zp^{\lastgenerator},\blindingfactor\in\Zp$ such that
\begin{equation*}
\hashsensors = \vectorgenerator^{\sensorvector}\cdot\hgenerator^{\blindingfactor}.
\end{equation*}
\begin{proc}(Consecutive difference) 
\label{proc:diff}
In this step the prover's goal is to compute the difference of consecutive values in the input vector, while keeping it hidden. Mainly, we want a provable value of 
    \begin{equation*}
    \hashsensors^d = \vectorgenerator^{\diffvector}\hgenerator^{\randomdiff}, 
    \end{equation*}
    with $\diffvector = [\sensorvector_1 - \sensorvector_2, \sensorvector_2-\sensorvector_3, \ldots, \sensorvector_{\sizevector-1} - \sensorvector_{\sizevector}, 0]$. 
    The intuition here is first to get a commitment of the iterated values of the sensor vector, then leverage the homomorphic property to subtract this commitment with $\hashsensors$ and finally provably replace the value in position $\sizevector$ by zero. To compute the iterated value, the prover first iterates the base generators, to get
\begin{equation*}
    \vectorgeneratoriter = [\generator_{\sizevector}, \generator_1, \ldots, \generator_{\sizevector-1}].
\end{equation*}     
Note that this step can be performed by the verifier as the generators are public. It then commits the sensor vector with this base
\begin{equation*}
 \hashsensoriter = \vectorgeneratoriter^{\sensorvector}\cdot\hgenerator^{\blindingiter},
\end{equation*}
with $\blindingiter\randin\Zp$, and generates a proof of equality, 
\begin{equation*}
\proofequality = \proofgen{\Pi_{Eq}}(\vectorgenerator, \vectorgeneratoriter, \hgenerator, \hashsensors, \hashsensoriter; \sensorvector, \blindingfactor, \blindingiter).
\end{equation*}
Note that 
\begin{equation*}
\hashsensoriter = \generator_1^{\sensorvector_2}\cdots\generator_{n-1}^{\sensorvector_{\sizevector}}\cdot\generator_n^{\sensorvector_1}\cdot\hgenerator^{\blindingiter},
\end{equation*}
so now we can simply subtract the two commitments to get 
    \begin{multline*}
         \overline{\hashsensors} = \hashsensors /  \hashsensoriter = 
        \generator_1^{v_1 - v_2}\generator_2^{v_2-v_3} \cdots \\ 
        \generator_{\sizevector-1}^{v_{\sizevector-1} - v_{\sizevector}}\generator_{\sizevector}^{v_{\sizevector} - v_1}\cdot\hgenerator^{\blindingfactor-\blindingiter}.
    \end{multline*}
    Finally, the prover replaces the value in the exponent of $\generator_{\sizevector}$ by a zero, to get the final commitment:
    \begin{equation}
    \diffcomm = \generator_1^{v_1 - v_2}\generator_2^{v_2-v_3} \cdots\generator_{\sizevector-1}^{v_{\sizevector-1} - v_{\sizevector}}\generator_{\sizevector}^0\cdot\hgenerator^{\randomdiff}, 
    \end{equation}
    and generates a proof of correctness, 
    \begin{equation}
    \proofzero=\proofgen{\Pi_0}(\vectorgenerator, \hgenerator, \overline{\hashsensors}, \diffcomm; \diffvector, \blindingfactor - \blindingiter, \blindingdiff).
    \end{equation} 
    It stores $\Delta = \left[\diffcomm, \hashsensoriter, \proofequality, \proofzero\right]$.
\end{proc}
\begin{proc}(Sum of vectors) 
\label{proc:sum}
The prover now computes $\addition = N\cdot \average$, mainly, the sum of all values. To provably compute this, we leverage \ipzkp between the initial commitment, $\hashsensors$, and a Pedersen commitment with base $\vectorhgenerator$ of the one vector, $\vectorhgenerator^\onevector$, to prove that a third commitment, $\avgcomm = \texttt{Comm}(\addition, \randomavg)$, commits to the sum of the committed values in $\hashsensors$, 
    \[\langle \sensorvector, \onevector\rangle = \addition.\]
    The user proves correctness of the commitment, 
    \begin{equation*}
    \proofipavg = \proofgen{\Pi_{IP}}(\vectorgenerator, \vectorhgenerator, \generator, \hgenerator, \hashsensors\cdot\vectorhgenerator^\onevector, \avgcomm; \sensorvector, \blindingfactor).
    \end{equation*} 
    It stores both values $\text{M} = \left[\avgcomm, \proofipavg\right]$.
    It repeats the same steps as above with the commitment of the consecutive difference vector, resulting in a commitment of the average, $\avgcommdiff$, and a proof of correctness, $\proofipavgdiff$. It stores both values $\text{M}' = \left[\avgcommdiff, \proofipavgdiff\right]$.
\end{proc}
\begin{proc}(Standard deviation) 
\label{proc:std}
To calculate a factor of the standard deviation, $\std$, we first compute the variance, $\variance$. Recall that 
    \[\variance = \frac{1}{N}\sum_{i=1}^N(v_i - \average)^2,\]
    or written differently
    \[\variance = \frac{1}{N}\langle \sensorvector - \boldsymbol{\average}, \sensorvector - \boldsymbol{\average}\rangle,\]
    where $\vectoraverage$ is a vector with $\average$ in all its positions.
    For this we need the average, but only have a provable commitment of the sum, $\addition$. Hence, instead of computing the variance, we compute $N^3 \cdot \variance$ by leveraging the inner product proof and the arithmetic properties of the commitment function.  
    The intuition is the following: if we multiply each entry of $\sensorvector$ by $N$, we can get the following relation.
    \begin{multline}
    \label{eq:ip-relation}
    \langle N\cdot \sensorvector - \vectoraddition, N\cdot \sensorvector - \vectoraddition\rangle = \\
    \langle N\cdot \sensorvector - N\cdot \vectoraverage, N\cdot \sensorvector - N\cdot \vectoraverage\rangle = \\ \langle N \cdot (\sensorvector - \vectoraverage), N\cdot (\sensorvector - \vectoraverage)\rangle = \\ N^2\langle \sensorvector - \vectoraverage, \sensorvector - \vectoraverage\rangle = N^3 \cdot \variance.
    \end{multline}
    However, we only have $\hashsensors$, and a provable commitment of $\addition$ (not of $\vectoraddition$). Moreover, we need a commitment of $\sensorvector$ and $\vectoraddition$ under both bases ($\vectorgenerator$ and $\vectorhgenerator$). To this end, the prover computes the following steps. First, it computes the commitment of $\vectoraddition$ with both bases. To this end, the prover first computes a product of all the bases, 
\begin{equation*}
\generator_{\Pi} = \prod_{i=1}^{\sizevector}\generator_i \hspace{2mm}\text{ and }\hspace{2mm} \hgenerator_{\Pi} = \prod_{i=1}^{\sizevector}\hgenerator_{i}.
\end{equation*}   
Note that this step is again reproducible by the verifier, and hence no proof is required. Next it commits the average using these products as a base, to get 

\begin{equation*}
\gcommitmentaddition = \generator_{\Pi}^{\addition}\cdot\hgenerator^{\randomgadd} = \generator_1^{\addition}\cdots\generator_{\sizevector}^{\addition}\cdot\hgenerator^{\randomgadd},
\end{equation*} 
and
\begin{equation*}
\hcommitmentaddition = \hgenerator_{\Pi}^{\addition}\cdot\hgenerator^{\randomhadd} = \hgenerator_1^{\addition}\cdots\hgenerator_{\sizevector}^{\addition}\cdot\hgenerator^{\randomhadd},
\end{equation*}
with $\randomgadd, \randomhadd\randin\Zp$.
It proves equality between the opening of  $\avgcomm, \gcommitmentaddition$ and $\hcommitmentaddition$ using $\proofequality$, and stores the two proofs, 
\begin{equation*}
    \proofequalityg = \proofgen{\Pi_{Eq}}(\generator_{\Pi}, \generator,  \hgenerator, \avgcomm, \gcommitmentaddition; \average, \randomavg, \randomgadd),
\end{equation*}
and 
\begin{equation*}
    \proofequalityh = \proofgen{\Pi_{Eq}}(\hgenerator_{\Pi}, \generator, \hgenerator, \avgcomm, \hcommitmentaddition; \average, \randomavg, \randomhadd),
\end{equation*} 
one for each base. Finally, the prover commits to $\sensorvector$ with randomness $\randomsensorhcomm$ with $\vectorhgenerator$ as bases, 
\begin{equation*}
\hcommitmentsensor = \vectorhgenerator^{\sensorvector}\hgenerator^{\randomsensorhcomm},
\end{equation*}
and proves equality of opening with respect to $\hashsensors$, getting 
\begin{equation*}
    \proofequalitys = \proofgen{\Pi_{Eq}}(\vectorgenerator, \vectorhgenerator, \hgenerator, \hashsensors, \hcommitmentsensor; \sensorvector, \blindingfactor, \randomsensorhcomm).
\end{equation*}  
This allows the prover to leverage relation~(\ref{eq:ip-relation}) to provably compute a commitment of a factor of the variance using the proof presented in Section~\ref{sec:zk-ip} To this end, it computes
\begin{multline*}
    \ipproductLR = \hashsensors^N/\gcommitmentaddition\cdot\hcommitmentsensor^N/\hcommitmentaddition = \\ \vectorgenerator^{N\cdot\sensorvector - \vectoraddition}\cdot\vectorhgenerator^{N\cdot\sensorvector - \vectoraddition}\cdot\hgenerator^{N\cdot\blindingfactor - \randomgadd + N\cdot\randomsensorhcomm - \randomhadd}
\end{multline*}
and the commitment of the factor of the variance, 
\begin{equation*}
    \commitmentvariance = \texttt{Comm}(N^3\cdot\variance, \randomvariance).
\end{equation*} 
It generates a proof of correctness 
\begin{multline*}
    \proofipvariance = \proofgen{\Pi_{IP}}(A_S, \commitmentvariance, \vectorgenerator, \vectorhgenerator, \generator, \hgenerator; \\ 
    \sensorvector, \addition, \blindingfactor, \randomgadd, \randomhadd, \randomsensorhcomm, \randomvariance).
\end{multline*}
    
    Finally, the prover needs to compute the square root of the variance. To this end, it commits to the floor of the square root of $N^3\cdot\variance$, $\commitmentstd = \commit(\lfloor\sqrt{N^3\cdot\variance}\rfloor, \randomsqrt)$.  Then, the prover leverages the square root proof introduced in Section~\ref{sec:crypto-background}, and generates
    \begin{equation*}
        \proofsqrt^{\std} = \proofgen{\proofsqrt}(\commitmentvariance, \commitmentstd, \generator, \hgenerator; \variance, \std, \randomvariance, \randomsqrt).
    \end{equation*}
This results in a provable commitment of a factor of the floor of the standard deviation, $\commitmentstd$. The prover stores, 
\begin{multline*}
\Lambda = \left[\gcommitmentaddition,\hcommitmentaddition,\hcommitmentsensor,\proofequalityg, \proofequalityh,\proofequalitys,\right. \\
\left.\commitmentvariance, \proofipvariance, \commitmentstd, \proofsqrtstd\right].
\end{multline*}
It repeats the same steps above with the consecutive difference vector (and average), resulting in
\begin{multline*}
\Lambda' = \left[\gcommitmentadditiondiff,\hcommitmentadditiondiff,\hcommitmentsensordiff,\proofequalitygdiff, \proofequalityhdiff,  \right.\\\left. \proofequalitysdiff,
\commitmentvariancediff, \proofipvariancediff, \commitmentstddiff, \proofsqrtstddiff\right].
\end{multline*}
\end{proc}

\begin{proc}(Computing SVM score)
\label{proc:score}
Provably computing SVM score (Eq.~\ref{eq:svm-score}) reduced to provably computing $\sum_{i=1}^{N} f_i \floor*{\frac{w_i}{S_i} 10^d}$ where $f_i$ are the features. However, note that we do not have the features themselves, but a factor of them. Hence, with this scheme, we need to compute instead $r = \sum_{i=1}^{N} f_i \floor*{\frac{w_i}{N_i \cdot S_i} 10^d}$ where $N_i$ equals $N$ if $f_i$ is an average, and $N^{3/2}$ if it is a standard deviation (note that we have different factors of each). Again, $\frac{w_i}{N_i \cdot S_i}$ is public. Let $\texttt{Comm}_i = \commit(f_i, r_i)$ be the commitment of feature $f_i$ with blinding factor $r_i$. The prover computes the following: 
\begin{multline}
\label{eq:linearcombinationscommitments}
\commitmentresult = \prod_{i=1}^{N}\texttt{Comm}_i^{\floor*{\frac{w_i}{N_i \cdot S_i} 10^d}} =\\
\prod_{i=1}^{N}\commit\left(f_i\cdot\floor*{\frac{w_i}{N_i \cdot S_i} 10^d}, r_i\cdot\floor*{\frac{w_i}{N_i \cdot S_i} 10^d}\right)  = \\
\commit\left(\sum_{i=1}^{N} f_i \floor*{\frac{w_i}{N_i \cdot S_i} 10^d}, \randomresult\right), 
\end{multline}
where $\randomresult=\sum_{i=1}^{N} r_i \floor*{\frac{w_i}{N_i \cdot S_i} 10^d}$, is the blinding factor known to the prover. 
Once these operations have been performed, the prover stores the opening of the commitment, $\openingscore = \sum_{i=1}^{N} f_i \floor*{\frac{w_i}{N_i \cdot S_i} 10^d}$, and the blinding factor $\randomresult$.
\end{proc}
\begin{proc}(Sending values)
\label{proc:send}
The prover sends to the verifier the following tuple
    \begin{equation*}
    \left[\hashsensors, \Delta, \text{M}, \text{M}', \Lambda, \Lambda',\openingscore, \randomresult\right]
    \end{equation*}
\vspace{5mm}
\end{proc}
%
%
\vspace{-0.5in}
\point{Proof Verification} \noindent The verifier then performs all respective linear combinations  commitments, and verifies the zero-knowledge proofs. If any proof fails or the evaluation of the model over $\openingscore$ fails, the verifier denies the request. Else, it accepts it. More precisely, the verifier follows the following steps: 
\begin{proc}(Verifying consecutive difference) 
\label{proc:verif-diff}
The verifier begins by iterating the base of generators:
\begin{equation*}
    \vectorgeneratoriter = [\generator_{\sizevector}, \generator_1, \ldots, \generator_{\sizevector-1}].
\end{equation*}    
\sloppy
and then verifies the proof of opening equality, 
\[
\proofequality.\texttt{Verif}(\vectorgenerator, \vectorgeneratoriter, \hgenerator, \hashsensors, \hashsensoriter)\stackrel{?}{=}\top.
\] 
Next it computes the subtraction commitment to get
\begin{equation*}
 \overline{\hashsensors} = \hashsensors / \hashsensoriter.
\end{equation*}
Finally, it verifies that the proof 
\[
\proofzero.\texttt{Verif}(\vectorgenerator, \hgenerator, \overline{\hashsensors}, \diffcomm)\stackrel{?}{=}\top.
\]
\end{proc}
\begin{proc}(Verifying sum of vectors)
\label{proc:verif-sum}
The verifier checks the inner product proof 
\[
\proofip^{\average}.\texttt{Verify}(\vectorgenerator, \vectorhgenerator, \generator, \hgenerator, \hashsensors\cdot\vectorhgenerator^\onevector, \avgcomm)\stackrel{?}{=}\top.
\]
It repeats this step for the consecutive difference commitment and proof. 

\end{proc}
\begin{proc}(Verifying standard deviation) 
\label{proc:verif-std}
The verifier first computes a product of all the bases, 
\begin{equation*}
\generator_{\Pi} = \prod_{i=1}^{\sizevector}\generator_i \hspace{2mm}\text{ and }\hspace{2mm} \hgenerator_{\Pi} = \prod_{i=1}^{\sizevector}\hgenerator_{i}.
\end{equation*}   
Next it verifies the proofs of equality of commitments using these bases
\[
\proofequality^G.\texttt{Verify}(\generator_{\Pi}, \generator,  \hgenerator, \avgcomm, \gcommitmentaddition)\stackrel{?}{=}\top,
\] 
and 
\[
\proofequality^H.\texttt{Verify}(\hgenerator_{\Pi}, \generator,  \hgenerator, \avgcomm, \hcommitmentaddition)\stackrel{?}{=}\top.
\]
Next, the verifier checks that $\hcommitmentsensor$ commits to the input vector
\[
\proofequality^{S}.\texttt{Verify}(\vectorgenerator, \vectorhgenerator, \hgenerator, \hashsensors, \hcommitmentsensor)\stackrel{?}{=}\top.
\]
Now the verifier needs to generate the commitments under which the inner product proof of the variance will verify against. To this end it computes
\begin{equation}
\label{eq:variance-relation}
L = \hashsensors / \gcommitmentaddition \hspace{2mm}\text{and}\hspace{2mm} R = \hcommitmentsensor / \hcommitmentaddition
\end{equation}
and uses them to verify the inner product proof
\[
\proofip^{\variance}.\texttt{Verify}(L\cdot R, \commitmentvariance, \vectorgenerator, \vectorhgenerator, \generator, \hgenerator)\stackrel{?}{=}\top.
\]
\noindent Finally, the verifier checks the correctness of the factor of the standard deviation commitment
\[
\proofsqrt^{\std}.\texttt{Verify}(\commitmentvariance, \commitmentstd, \generator, \hgenerator)\stackrel{?}{=}\top.
\]
It repeats these steps for the consecutive difference commitments.

\end{proc}
\begin{proc}(Computing SVM score)
\label{proc:verif-score}
Finally, the verifier computes the same linear combinations as the prover, 
\begin{equation}
\label{eq:commit-result}
\commitmentresult' = \prod_{i=1}^{N}\texttt{Comm}_i^{\floor*{\frac{w_i}{N_i \cdot S_i} 10^d}}, 
\end{equation}
and checks the validity of the received opening, 
\[
\opening(\commitmentresult',\openingscore, \randomresult, \generator, \hgenerator)\stackrel{?}{=}\top.
\]
It uses this value to compute the final score as described in Equation~(\ref{eq:svm-score}). If any of the checks fail or the score determines the user is a bot, it returns $\bot$, otherwise it returns $\top$. 
\end{proc}
%
By extending the inner product proof presented in~\cite{bulletproofs} to a zero-knowledge proof and leveraging the arithmetic properties of Pedersen commitments, we present \innerProdName, a privacy-preserving SVM evaluation model. 

\section{System Implementation}
\label{sec:implementation}
To assess the feasibility and effectiveness of our approach we developed (i) an open source library of zkSVM, and (ii) a prototype Android SDK of \name. 

\subsection{Enclosing SVM Result in a ZKP}

\point{The \innerProdName library}
To prove the integrity of the classification results, we developed an open-sourced Rust library that implements the logic presented in Protocol~2, on enclosing SVM results in zero-knowledge proofs. To this end, we additionally implement the Pedersen Commitment ZKPs as described in Section~\ref{sec:crypto-background}. 
For the proofs $\proofsquare,\proofequality, \proofzero$ and $\proofopening$, we base our implementation in the work presented in~\cite{veccommitments}. We use the range proof presented in~\cite{bulletproofs} and implemented in~\cite{bulletproofsimpl}. Finally, for $\proofip$, we implement the zero-knowledge proof presented in Section~\ref{sec:zk-ip} All the above proofs are implemented using the ristretto255 prime order group over Curve25519 by leveraging the curve25519-dalek~\cite{curve25519-dalek} library. All our proofs leverage the Fiat-Shamir heuristic to make the ZKPs non-interactive, which maintains all security properties under the Random Oracle model~\cite{ROM}.
To integrate this library with our detection engine, we use the Android NDK development kit.


\point{General-purpose \snark}
To compare the performance of \innerProdName, we implement 
the SVM execution using the \zokrates general-purpose \snark toolbox~\cite{zokrates}. \zokrates works as a high-level abstraction for the encoding of the computation to be proved into a \snark. \zokrates  constructs the ZKP by using the Rust implementation of Bellman's \cite{bellman} Groth16 \snark \cite{Groth16}. This construction has optimal proof size and verification time. However, this comes by trading off prover's computational complexity and the requirement of a trusted setup.



\subsection{\NAME Prototype}
We implemented a prototype SDK of \name for Android ($\sim$1,000 lines of Java). The SDK monitors Android's accelerometer and gyroscope during a touch event and, by applying a \mbox{pre-trained} model, it determines if it was performed by a human or not. For demonstration purposes, we created a demo app with a user interface that shows the output of the detection model and we provide publicly a video\footnote{\name demo: \url{https://youtu.be/U-tZKrGb8L0}} that demonstrates its functionality. In this demo, we test multiple scenarios to showcase the accuracy of our system: 

 \begin{figure*}[t]
    \centering
    \begin{minipage}{0.32\linewidth}
        \includegraphics[width=1.05\columnwidth]{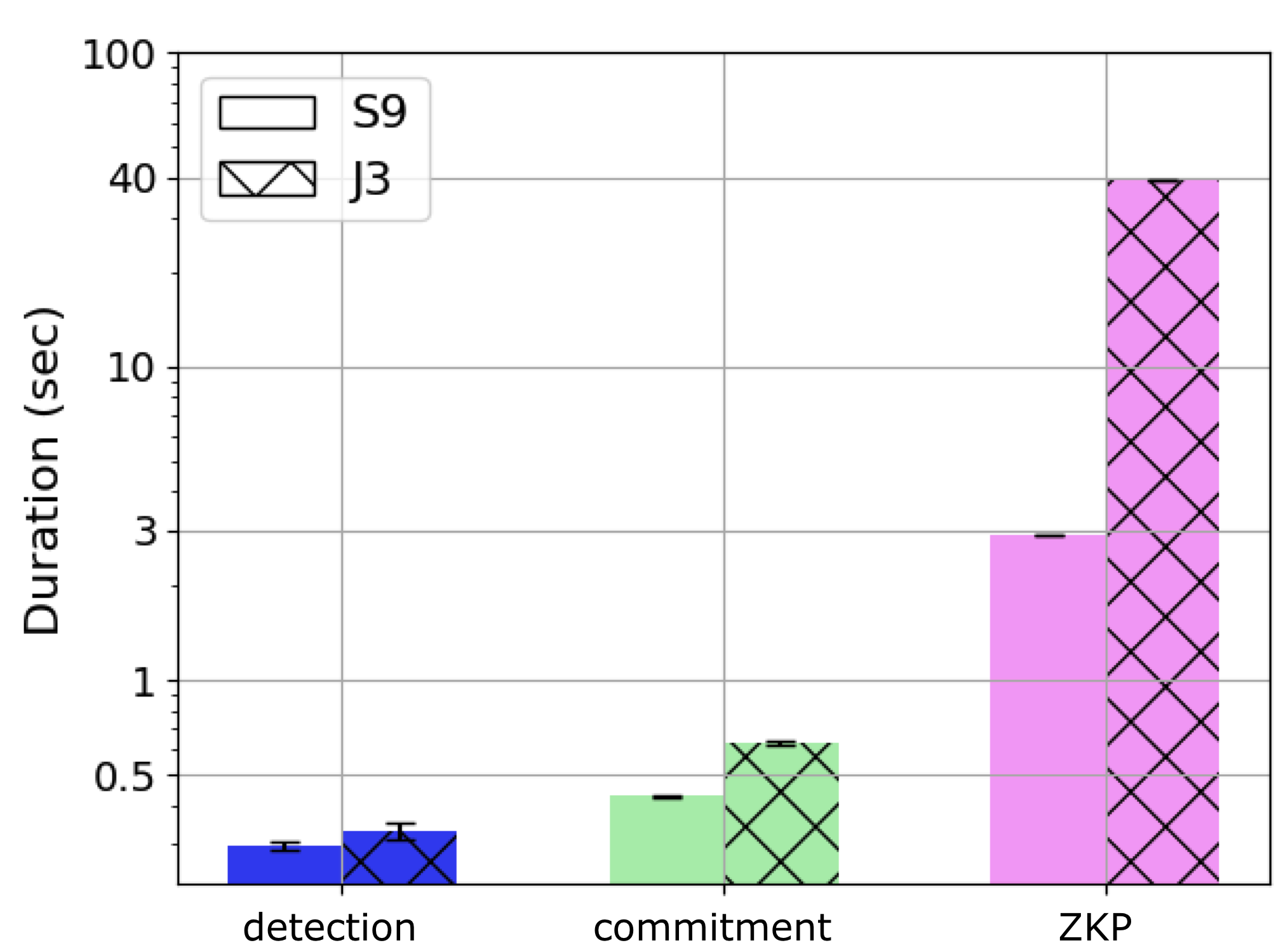}
        \caption{Execution time per operation. On commodity hardware (S9) humanness detection and commitments are extremely fast (\ie about \latencySdetection), when the ZKP generation lasts about \latencySzkp  seconds.}
        \label{fig:duration}
    \end{minipage}
    \hfill
    \begin{minipage}{0.32\linewidth}
        \centering
        \includegraphics[width=1.05\columnwidth]{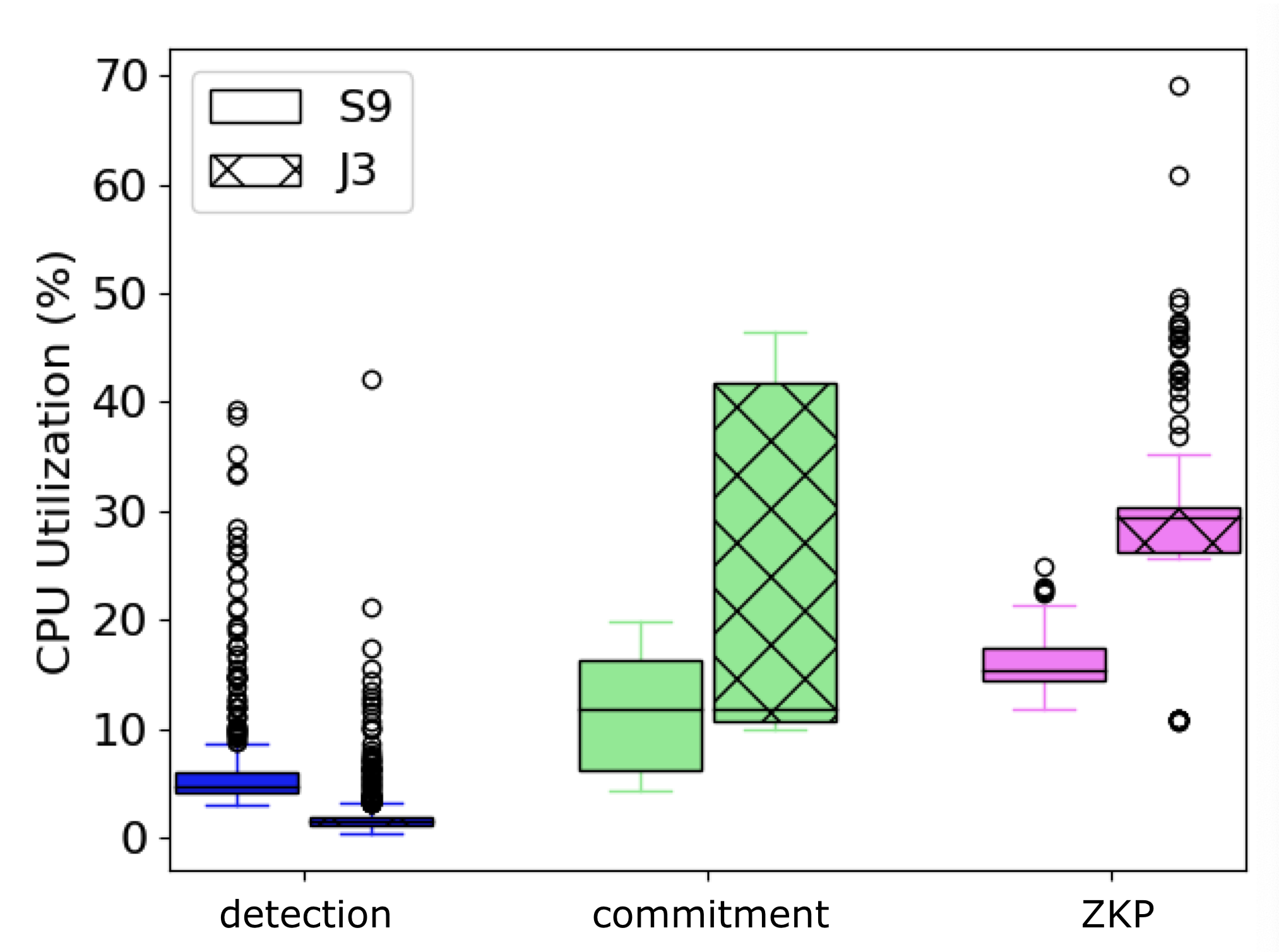}
        \caption{CPU utilization per operation. On commodity hardware (S9), the ZKP generation is the most expensive operation, with a median CPU consumption of about~15\% on commodity hardware.}
        \label{fig:cpu}
    \end{minipage}
    \hfill
    \begin{minipage}{0.32\linewidth}
        \includegraphics[width=1.05\columnwidth]{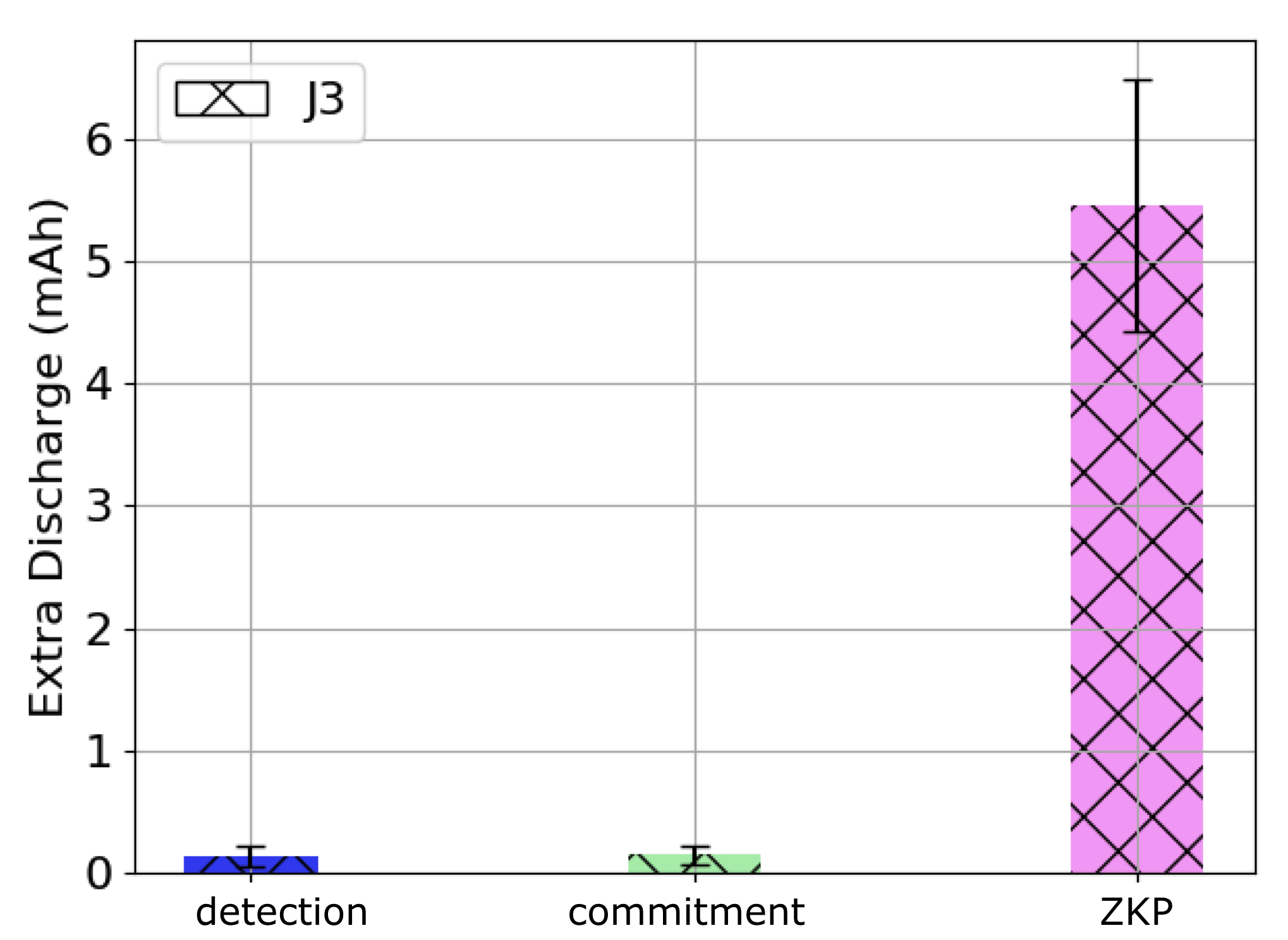}
        \caption{Energy consumption per operation on a low-end hardware. Energy consumption for detection and committing operation is negligible, when for the ZKP computation \name consumes about~5 mAh or~0.2\% of the device's battery.}
        \label{fig:battery}
    \end{minipage}
 \end{figure*}

\begin{enumerate}[topsep=0.2cm]
    \item The device is resting on a steady platform and: 
    \begin{enumerate}
    \item A human is performing clicks.
    \item Clicks are simulated.
    \end{enumerate}
    \item The device is docked on a swing motion device that produce artificial movement. 
    \item The device is held in one hand and:
      \begin{enumerate}
        \item A human is performing clicks.
        \item Clicks are simulated.
    \end{enumerate}  
\end{enumerate}

After reading the output of the accelerometer and gyroscope sensors, the \name SDK applies, on the background, our pre-trained model and classifies the origin of the touch-screen event (\ie performed by a human or not). For the generation of the ZKPs and the model's evaluation, we leverage the library we implemented and described previously, which we call from the mobile device using the sensor data.
The pre-trained model, is generated on a server of ours. Apart from generating and distributing the trained model, the server also acts as the external auditor that verifies the validity of the transmitted attestation results.

\section{Performance Evaluation}
\label{sec:evaluation}



In this section, we set out to explore the performance of humanness attestation in \name. More specifically, we benchmark our Android prototype with respect to  the \emph{duration} of its main operations: (i) humanness classification, (ii) Pedersen commitment computation, and (iii) zero-knowledge proof construction. Next, we  evaluate  general resource utilization metrics: (a) CPU, (b) memory, and (c) battery consumption. Our tests cover the three key operations of a humanness attestation in \name, and a comparative baseline: 

\begin{enumerate}[leftmargin=0.6cm,topsep=0.2cm]
    \item The \emph{baseline}, where we run our demo application which uses \name service (see Section~\ref{sec:implementation}) and several artificial clicks are generated. 
    \item The \emph{detection} operation, where sensors input is collected and humanness classification realized on \name. 
    \item The \emph{commitment} operation where the Pedersen commitment computation is taking place.
    \item The \emph{ZKP} operation where the proof of correct attestation is constructed. 
\end{enumerate}
We test and compare the different implementations described in Section~\ref{sec:implementation}: (i) the general-purpose \snark and (ii) our \innerProdName. Note that for \snark we ignore the time elapsed while computing the trusted setup, as this cannot be computed by the client. We run each stage for an hour and we ensure the same number of artificial clicks by using as an interval the duration of the longest operation (ZKP) as empirically measured on each device under test (Figure~\ref{fig:duration}). 

\point{Setup} We leverage a testbed composed of two Android devices representative of a mid-end (Samsung Galaxy S9, model 2018) and a low-end (Samsung Galaxy J3, model 2016) device to inspect what is the worst performance a user can get on a cheap (around 90 USD) device. The S9 mounts an octa-core processor (a Quad-Core Mongoose M3 at~2.7GHz and a Quad-Core ARM Cortex-A55 at~1.8Ghz), while the J3 is equipped with a quad-core ARM Cortex A53 at~1.2 Ghz. The S9 also has twice as much memory (4 GB when J3 has~2 GB) and a larger battery (3,000 mAh when the battery of J3 is~2,600 mAh).  The low-end device (J3) is part of Batterylab~\cite{batterylab2019,batterylab}, a distributed platform for battery measurements. It follows that fine grained battery measurements (via a Monsoon High Voltage Power Monitor~\cite{monsoon} directly connected to the device's battery) are available for this device. Automation of the above operations is realized via {\tt adb} run over WiFi to avoid noise in the power measurements caused by USB powering.

\point{Execution time}
Figure~\ref{fig:duration} shows the average duration (standard deviation as error-bars) of each \name's operation, per device, when considering \innerProdName. Regardless of the device, humanness classification and commitments are extremely fast, \ie about \latencySdetection and \latencyLhashing seconds even on the less powerful J3.  The ZKP generation is instead more challenging, lasting about \latencySzkp and \latencyLzkp seconds on the S9 and J3, respectively. 

To improve visibility, the figure omits results from the general purpose \snark solution. In this case, we measure commitment operations comprised between 24 and 190 seconds, and ZKP generation comprised between 175 and \zokratestime seconds, depending on the device. This suggests one order of magnitude speedup of \innerProdName versus the more generic \snark solution. Given the significant difference between \innerProdName and \snark, in the following we mostly focus on results obtained via \innerProdName. 

For the verification time, the general purpose \snark (\verifiertimesnark) outperforms \innerProdName(\verifiertimeadhoc). This is expected as the Groth16 approach used by \zokrates has a big prover overhead and a trusted setup in exchange of minimal communication and verification time overhead.   
However, in \name's scenario, these verification times can be handled by the server, and instead, \innerProdName makes the prover times reasonable (as shown, the entire attestation takes a bit less than \proverLatencyS seconds) and removes the need for trusted setup.

\point{CPU and memory utilization}
Figure~\ref{fig:cpu} shows CPU usage per operation and device. Since no significant difference was observed between baseline and detection operation, we improve the figure visibility by reporting only one boxplot representative of both operations. The figure shows minimal CPU utilization associated with humanness classification and commitment operations. The counter-intuitive higher CPU usage at the S9  is due to the fact that this is a personal device with potential background activities from other apps. Even on the less powerful J3, committing only consumes about 12\% of CPU (median value across devices) with peaks up to~45\% on the J3. The ZKP generation is the most expensive operation, showing a median CPU consumption of about~15\% and~30\%, on respectively the S9 and J3. Overall, the CPU analysis suggests minimal impact of \name's operation and feasibility even on entry-level devices like the J3. 

In our tests, we also collected memory usage of \name's via \texttt{procstats}. Detailed results are omitted since \name's memory consumption is negligible, \ie less than 20MB regardless of device and operation. In comparison, the \snark solution requires up to 1GB of memory due to the data generated during the trusted setup. 
This is quite limiting in presence of low-end devices which might not have that amount of free memory, requiring swapping and thus a further increase in execution time.

\point{Battery consumption}
Figure~\ref{fig:battery} reports the battery discharge (in mAh) associated with \name's key operations ( detection, commitment, ZKP creation) for the J3 -- given the S9 is a personal device, we were unable to wire its battery with the power meter. As expected from the previous results, the battery overhead imposed by  \name's detection and committing operation is negligible. Further, ZKP computation only consumes about~5 mAh or~0.2\% of the J3's battery~(2,600 mAh). Even assuming one \name's humanness verification every hour of device usage, this would amount to under 1\% of battery discharge for the average user (3h15min on average, with only 20\% of users using the device more than 4h30min~\cite{screentime}). 

\point{Bandwidth consumption}
Finally, we compute what is the bandwidth consumption required by the user to send a proof. The proof consists of the commitments of the difference vector, the average and standard deviation, for each of the input vectors the user submits. We use data from two sensors, namely the gyroscope and accelerometer, and for each sensor, we use three axis data. Moreover, we split each period into two segments as explained in Section~\ref{sec:classification}. This results in a total of 12 input vectors. For every input vector, the proof consists of \singleProof (83 compressed points and 354 scalars). In our library, we implemented the trivial construction, where we build 12 of such proofs in parallel, resulting in \fullNonOptimisedProof. However, there are ways to reduce this overhead, for which we provide the estimates. The opening and equality proofs we used (introduced in Section~\ref{sec:crypto-background}) could be improved, by implementing them using the techniques presented in~\cite{BCC+16}. This would reduce the complexity from linear to logarithmic, resulting in x3 improvement, with a size of the full proof of~\estimateEqOpimprovement. Finally, we could use the batching techniques for the range proofs, as presented in~\cite{bulletproofs}. This would further reduce the size of the proof to~\estimateRangeBatchImprovement. 

\subsection{\NAME Vs. reCAPTCHA}
 
As a next step, we compare the performance of \name with the state-of-the-art privacy-preserving humanness attestation mechanism (\ie visual CAPTCHA). To do so, we developed an Android app which embeds  reCAPTCHA for Android~\cite{android_captcha}. The app is minimal\footnote{Source code: \url{https://github.com/svarvel/CaptchaTest}} to ensure its performance evaluation covers only the CAPTCHA aspect rather than any extra components. For the same reason, we opted for  Android reCAPTCHA rather than setting up a webpage with a  CAPTCHA to solve. This alternative approach would require an Android browser for testing and the performance evaluation would be tainted by the extra cost of running a full browser. We did not evaluate Privacy Pass~\cite{privacyPassIETF} for two reasons: 1) it currently requires a full browser along with an add-on, 2) it lacks support on mobile devices. Note that we do not expect critical performance difference between Privacy Pass and Android reCAPTCHA since they use a very similar strategy. Their difference instead lies in how invasive they are, both in how frequently they require user input and from a sensor data collection standpoint. 

Using the above application we setup the following experiment. We enable remote access to the S9 device via the browser using Android screen mirroring~\cite{scrcpy} coupled with noVNC~\cite{novnc}. Then we asked~10 volunteers to visit the device from their browsers and solve CAPTCHAs as needed by the app. The app was coded such that users can continuously request for a new CAPTCHA to solve. Note that Android reCAPTCHA, as reCAPTCHA v3, leverages client side behavior to minimize friction, \ie whether to ask or not a user to solve a visual CAPTCHA like ``click on all images containing a boat''. It follows that often users do not need to solve any visual CAPTCHA.  We label \textit{automatic} all the samples we collected where our volunteers did not have to solve a CAPTCHA. We instead label \textit{manual} all the samples where human interaction was needed. To increase the chance of showing an actual CAPTCHA, we created 15 CAPTCHAs that the app rotates on. 


Over one week, our volunteers have requested about~500 CAPTCHAs with a~70/30 split:~350 automatic and 150 manual. Table~\ref{tab:comparison} directly compares \name with Android reCAPTCHA with respect to: execution time, CPU and memory utilization, and bandwidth consumption. The median was reported for each metric, further differentiating between automatic and manual for Android reCAPTCHA. The table shows that \name adds about~1.6 seconds to the time required by Android reCAPTCHA when no user interaction is needed. 
\name instead saves a whole 6 seconds to the (median) user by never requiring any interaction. Even considering the fastest user in our experiment (5.4 seconds), \name is about 2x faster -- and 10x faster than the slowest user (28.5 seconds). This is possible because \name removes the need of user interaction, at the cost of a higher risk for replay attacks. With respect to CPU and memory utilization, Table~\ref{tab:comparison} shows that the two mechanisms are quite similar and both very lightweight. Bandwidth-wise, reCAPTCHA outperforms \name (16 versus 160KB), which communication overhead is still minimal and bearable even by devices with very little connectivity. 

 \begin{table}[t]
     \centering
     \begin{tabular}{lrr}
     \toprule
        & {\bf \name} & {\bf reCAPTCHA} \\
        & & {\bf Automatic/Manual} \\
        \midrule 
        Overall time    & \proverLatencyS & 1.4/8.9 sec \\
        CPU utilization    & 15\%                    & 3/15\%   \\
        Memory utilization & 20 MB                   & 20 MB    \\
        Replay protection & No & Yes \\
        Consumed bandwidth & \fullNonOptimisedProof  & 16 KB     \\
        \bottomrule
     \end{tabular}
     \caption{Performance of \name vs. Android reCAPTCHA.}\vspace{-0.3in}
     \label{tab:comparison}
 \end{table}

Table~\ref{tab:comparison} currently reports on "overall time" which includes both the computation time and the time required to report to the server. While for Google we have no control on the server endpoint -- located within 10ms in our experiments -- in our experiments the server runs in the same LAN with 1-2 ms delay (negligible). We currently use HTTP (POST) + TLS1.3 to return the proof. For TCP, we use an unmodified kernel running Cubic with an initial window of 10packets (1500B MTU). Given our proof has a size of 160KB, the content delivery requires a worst case of: 1 RTT (for TCP) + 1 RTT (for TLS, in case of unknown server) + ~3 RTT for TCP to transfer the data -- assuming slow start (doubling of cwnd), aka 15K (10 MTU sized packets) + 30K + 60K + 60K (1/2 of the last cwnd available). This sums up to about 3/4 RTTs, which we could further reduce assuming a larger initial cwnd, or using QUIC -- thus bringing the duration down to a maximum of 2RTTs. Assuming a CDN runs such a service, as Google does, this would thus cost between 20ms (with optimization) up to 100ms. Assuming a very bad client connection, e.g., on mobile with RTT of 150ms, then this would cost an extra 300 to 750ms.

\subsection{Summary}
Our experiments show that the general purpose \snark is not a viable solution for mobile-based ZKP computation (\zokratestime sec and \zokratesbatt battery drain on a low-end device). By designing our own model (\ie \innerProdName), we reduce ZKP's execution time by~$10\times$, achieving a duration of a bit less than \proverLatencyS  sec and  \proverBatteryPerc battery drain. This execution time is comparable with today's visual CAPTCHA solving time, 9.8 sec on average~\cite{captchaUX}), thus making \name a serious competitor to state-of-the-art mechanisms for humanness attestation. 
\section{Related Work}

\point{Assessing Humanness}
To prevent automated programs from abusing online services, the widely adopted solution is to deploy a CAPTCHA system. 
However, text-based CAPTCHA schemes have been proven to be insecure as machines achieved~99.8\% success rate in identifying distorted text~\cite{yan2008low,5395072,goodfellow2013multi}. Audio-based CAPTCHAs have also been used to assist visually impaired people, but they are difficult to solve, with over half of users failed during their first attempt~\cite{tasidou2012user}. Therefore, CAPTCHA service providers started to test image-based CAPTCHA schemes, which require users to select images that match given description~\cite{Google2014}. 
Nevertheless, in~\cite{sivakorn2016robot,Zhou:2018:BGR:3280489.3280510} authors demonstrated that more than 70\% of image-based Google and Facebook CAPTCHAs can be efficiently solved using deep learning.

In~\cite{walgampaya2010real}, authors designed a multi-level data fusion algorithm, which combines scores from individual clicks to generate a robust human evidence. These CAPTCHA systems require users to perform additional tasks worsening user experience, especially on mobile devices~\cite{reynaga2013usability}. ReCAPTCHA v2 uses a risk analysis engine to avoid interrupting users unnecessarily~\cite{Google2019}. This engine collects and analyses relevant data during click events. ReCAPTCHA v3 no longer requires users to click but instead it studies user interactions within a webpage and gives a score that represents the likelihood that a user is a human~\cite{Google2018}. Although these CAPTCHA schemes are invisible to users, a plethora of sensitive data, including cookies, browser plugins, and JavaScript objects, is collected~\cite{LaraOReilly2015} that could be used to fingerprint the user. 

With the proliferation of smartphones, various approaches leverage the variety of available sensors. Most of them require users to perform additional motion tasks. In~\cite{10.1007/978-3-319-02937-5_11}, authors showed that waving gestures could be used to attest the intention of users. In~\cite{guerar2015completely}, authors designed a bot detection system that asks users to tilt their device according to the description to prove they are human. In~\cite{10.1007/978-3-319-45572-3_3}, authors presented a movement-based CAPTCHA scheme that requires users to perform certain gestures (\eg hammering and fishing) using their device. 
In~\cite{DeLuca:2012:TMO:2207676.2208544}, authors exploited touch screen data during screen unlocking to authenticate users. In~\cite{guerar2016using}, authors suggested a brightness-based bot prevention mechanism that generates a sequence of circles with different brightness when typing a PIN; users will input misleading lie digits in circles with low brightness. In~\cite{buriro2017evaluation}, authors proposed a behavioural-based authentication scheme, which uses timing and device motion information during password typing.

The work that is most closely related to ours is the Invisible CAPPCHA~\cite{guerar2018}. Similar to \name, authors leveraged the different device acceleration appearing on a finger touch and a software touch to make a decision about whether a user is a bot. 
However, Invisible CAPPCHA is not fully implemented as it requires a secure execution environment and its accuracy is low when device is stable on a table.
In addition, it only considers simple tap and vibration events; its accuracy on more complicated touch events (\eg drag, long press, and double tap) is unclear
In comparison, \name considers any types of touch events and works regardless of the device movement. To improve the accuracy, \name uses more data sources in addition to accelerometer and introduces context into the detection.

\point{Privacy-Preserving and Provable ML}
A potential approach to offer privacy-preserving machine learning is to evaluate the model locally, without sending data to a server. However, if, unlike \name, such approach is taken without proving correct evaluation of the model, then verification may be lost~\cite{DBLP:journals/corr/abs-1710-03275,Bilenko:2011:PCP:2020408.2020475,Guha:2011:PPP:1972457.1972475}. In cases such as bot detection the user's interest might be of faking the evaluation model, and this may be vulnerable to user attacks. 
To the best of our knowledge, there are only 2 papers aiming to provide provable machine learning  local evaluation without a trusted execution environment. The first one~\cite{Davidson:2014:MMO:2664243.2664266} tries to solve a similar problem, where personalization of a user device is done by evaluating a model locally on the user's machine. This work uses Bayesian classification, for which they need from 100-300 feature words. The generation of correct model evaluation for such range of feature words ranges from 30 to 80 seconds. Moreover, this study uses standard techniques for constructing zero-knowledge proofs, which give a big overhead to the verifier. For our particular use case (where the verifier may need to handle several requests simultaneously), such an overhead for the verifier is not acceptable. The second one~\cite{Danezis:2012:PCP:2359015.2359018} proposes a solution where after the evaluation of Random Forest and Hidden Markov models, the user generates a zero-knowledge proof of correct evaluation. However, this paper misses an evaluation study or availability of the code, which makes a study of the scalability of their approach inaccessible.

\section{Conclusion}
Service providers need a reliable way to attest whether a client is human or not and thus prevent user-side automation from abusing their services. Current solutions require (i) either additional user actions (\eg sporadically solve mathematical quizzes or pattern recognition) like CAPTCHAs or (ii) user behavioral data to be sent to the server, thus raising significant privacy concerns. 

In this paper, we propose \name: a novel continuous human attestation scheme, which is both (i) privacy-preserving and (ii) friction-less for the user. By leveraging the motion sensor outputs during touch screen  events, \name performs human attestation at the edge, on the user’s very own device, thus avoiding to transmit any sensitive information to a remote server. By enclosing the classification result in zero-knowledge proofs, \name guarantees the integrity of the attestation procedure. 
We tested our system under different attack scenarios: (i) when device is resting (on a table), (ii) when there is artificial movement from device's vibration, (iii) when the device is docked on a artificial movement generating swinging cradle, and it was able to detect if an action was triggered by a human with \modelaccuracy accuracy.
Performance evaluation of our Android prototype demonstrates that each attestation takes around \proverLatencyS seconds, with minimal impact on both CPU and battery consumption. 

\bibliographystyle{unsrt}
\bibliography{paper}

\section*{Appendix}
\appendix
\section{Sub-linear Inner Product Zero-Knowledge Proof}
\label{appendix:proof}

In this section, we prove that the zero-knowledge proof presented in section \ref{sec:crypto-background} provides perfect completeness, special honest verifier zero-knowledge and knowledge soundness. Firstly, we formally define these properties, which we reproduce as defined in~\cite{BCC+16}. 

\newcommand{\generatorproof}{\mathcal{K}}
\newcommand{\simulator}{S}
\newcommand{\adv}{\mathcal{A}}
\newcommand{\relation}{\mathcal{R}}
\newcommand{\extractor}{\mathcal{E}}
\newcommand{\commoninput}{pp}
\newcommand{\instance}{u}
\newcommand{\witness}{w}
\newcommand{\setrelation}{\mathcal{L}_{\relation}}

A zero-knowledge proof system is defined by three probabilistic polynomial time algorithms, $(\generatorproof, \prover, \verifier)$, the generator, prover and verifier. The generator takes as input a security parameter written in unary form, $1^{\secparam}$, and builds the common input of a proof, $\commoninput\leftarrow\generatorproof(1^{\secparam})$. In our paper, we use only common inputs that do not need to be honestly generated, meaning that the output of $\generatorproof$ can be publicly verified. The $\prover$ and $\verifier$ algorithms take as input $(\commoninput, \instance, \witness)$ and $(\commoninput, \instance)$ respectively. We denote the interaction between prover and verifier, and the latter's output (0 if valid or 1 otherwise) by $\langle\prover(\commoninput, \instance, \witness) || \verifier(\commoninput, \instance)\rangle$. We consider relations $\relation$ that consist of a three element tuple $(\commoninput, \instance, \witness)$, which we refer as the common input, instance and witness respectively. We define the set of all valid instances as $\setrelation = \{(\commoninput, \instance) | \exists \witness: (\commoninput, \instance, \witness)\in\relation\}$. The protocol $(\generatorproof, \prover, \verifier)$ is called a zero-knowledge proof system if it has perfect completeness, knowledge soundness and special honest-verifier zero-knowledge as described below. First, we introduce the notion of negligible function.
\begin{definition}[Negligible function]
A non-negative function $f:\mathbb{N}\rightarrow\mathbb{R}$ is called negligible if for every $\gamma\in\mathbb{N}$ there exists a $k_0\in\mathbb{N}$ such that for all $k\geq k_0, f(k)\leq 1/ k\gamma$.
\end{definition}
We now proceed with a formal definition of the properties a proof system needs to have to be considered a zero-knowledge proof. 
\begin{definition}[Perfect Completeness]
A proof system is perfectly complete if for all PPT adversaries $\adv$
\[
\Pr\Bigg[\
\begin{aligned}
\commoninput \leftarrow \generatorproof(); (\instance, \witness)\leftarrow \adv(\commoninput): \\ 
(\commoninput, \instance, \witness)\notin\relation \vee \langle\prover(\commoninput, \instance, \witness) || \verifier(\commoninput, \instance)\rangle = 1
\end{aligned}
\Bigg] = 1
\]
\end{definition}
Paraphrasing, this means that whenever prover and verifier proceed with the protocol, the verifier will always validate the proof. 
\newcommand{\state}{s}
\begin{definition}[Knowledge soundness]
A proof system has (strong black-box) computational knowledge soundness if for all DPT $\prover^*$ there exists a PPT extractor $\extractor$ such that for all PPT adversaries $\adv$
\[
\Pr\left[\
\begin{aligned}
\commoninput\leftarrow\generatorproof(1^\secparam); (\instance, \state)\leftarrow\adv(\commoninput);\\
\witness\leftarrow\extractor^{\langle\prover^*(\state) || \verifier(\commoninput, \instance)\rangle}(\commoninput, \instance): \\
b = 1 \wedge (\commoninput, \instance, \witness)\notin\relation
\end{aligned}
\right]
\]
is negligible with respect to $\secparam$.
\end{definition}
Here the oracle $\langle\prover^*(\state) || \verifier(\commoninput, \instance)\rangle$ runs a full protocol execution from the state $\state$, and if the
proof is successful it returns a transcript of the prover’s communication. The extractor $\extractor$ can ask the oracle to rewind the proof to any point and execute the proof again from this point on with fresh
challenges from the verifier. We define $b \in {0, 1}$ to be the verifier’s
output in the first oracle execution, \ie whether it accepts or not, and we think
of $\state$ as the state of the prover. 
If the definition holds also for unbounded $\prover^*$ and $\adv$ we say the proof has
statistical knowledge soundness.
The definition can then be paraphrased as saying
that if the prover in state $s$ makes a convincing proof, then we can extract a
witness.

\begin{definition}[Special Honest-Verifier Zero-Knowledge]
A proof system is computationally special honest-verifier zero-knowledge (SHVZK) if there exists a PPT simulator $\simulator$ such that for all state-full interactive PPT adversaries $\adv$ that output $(\instance, \witness)$ such that $(\commoninput, \instance, \witness)\in\relation$ and randomness $\phi$ for the verifier
\begin{multline*}
\Pr\left[\
\begin{aligned}
\commoninput\leftarrow\generatorproof(1^\secparam); (\instance, \witness, \phi)\leftarrow\adv(\commoninput); \\ 
\mathtt{view}_{\verifier}\leftarrow \langle\prover(\commoninput, \instance, \witness) || \verifier(\commoninput, \instance, \phi)\rangle: 
\\
\adv(\mathtt{view}_{\verifier}) = 1
\end{aligned}
\right] - \\
\Pr\left[\
\begin{aligned}
\commoninput\leftarrow\generatorproof(1^\secparam);(\instance, \witness, \phi)\leftarrow\adv(\commoninput); \\
\mathtt{view}_{\verifier}\leftarrow \simulator(\commoninput, \instance, \phi): \\
\adv(\mathtt{view}_{\verifier}) = 1
\end{aligned}
\right]
\end{multline*}
is negligible with respect to $\secparam$. We say the proof is statistically SHVZK if the definition holds against unbounded adversaries, and perfect SHVZK if the probabilities are equal.
\end{definition}
This definition can be paraphrased as saying that for every valid protocol run, a simulator can generate simulated random view which is indistinguishable from the original run. 
Having formalised the properties, we proceed with the proof of Theorem~\ref{thm:ip-zkp}:
\vspace{2mm}

\noindent\textit{Proof.}
We follow the lines of the proof presented in the work of Bünz \etal \cite{bulletproofs} to complete our proof. 
Completeness is trivial. To prove special honest verifier zero-knowledge we construct a simulator that generates valid statements which are indistinguishable from random. To this end, the simulator acts as follows: 
\begin{align*}
\challenge &\randin\Zp^* \\
\evaluatedleftpolynomial, \evaluatedrightpolynomial &\randin \Zp^{\lastgenerator} \\
\evaluatedinnerproductpolynomial, \blindingevalippoly, \blindingannouncement &\randin\Zp \\
\commitmentcoefficients_2 &\randin \group \\
\commitmentcoefficients_1 &= \left(\generator^{\evaluatedinnerproductpolynomial}\hgenerator^{\blindingevalippoly} \cdot 
\commitmentinnerproduct^{-1}\cdot\commitmentcoefficients_2^{-\challenge^2}\right)^{1/\challenge} \\
\commitmentblinders &= \left(\hgenerator^{\blindingannouncement}\cdot\vectorgenerator^{\evaluatedleftpolynomial}\cdot\vectorhgenerator^{\evaluatedrightpolynomial} \cdot \commitmentvectors^{-1}\right)^{1/\challenge}
\end{align*}

\noindent We can see that the simulated transcript,
\begin{equation*}
(\commitmentblinders, \commitmentcoefficients_1, \commitmentcoefficients_2; \challenge; \evaluatedleftpolynomial, \evaluatedrightpolynomial, \evaluatedinnerproductpolynomial, \blindingevalippoly, \blindingannouncement),
\end{equation*}
all elements except for $\commitmentblinders$ and $\commitmentcoefficients_1$ are random, and the latter two are computationally indistinguishable from random due to the DDH assumption. Also, note that $\commitmentblinders$ and $\commitmentcoefficients_1$ are generated following the verification equations, hence this simulated conversation is valid. 

Next, we construct an extractor to prove knowledge soundness. The extractor runs the prover with three different values of the challenge $\challenge$, ending with the following valid proof transcript:
\begin{align*}
& (\commitmentblinders, \commitmentcoefficients_1, \commitmentcoefficients_2; \challenge'; \evaluatedleftpolynomial', \evaluatedrightpolynomial', \evaluatedinnerproductpolynomial', \blindingevalippoly', \blindingannouncement') \\
& (\commitmentblinders, \commitmentcoefficients_1, \commitmentcoefficients_2; \challenge''; \evaluatedleftpolynomial'', \evaluatedrightpolynomial'', \evaluatedinnerproductpolynomial'', \blindingevalippoly'', \blindingannouncement'') \\
& (\commitmentblinders, \commitmentcoefficients_1, \commitmentcoefficients_2; \challenge'''; \evaluatedleftpolynomial''', \evaluatedrightpolynomial''', \evaluatedinnerproductpolynomial''', \blindingevalippoly''', \blindingannouncement''').
\end{align*} 

Additionally, the extractor invokes the extractor of the inner product argument. This extractor is proved to exist in the original paper of Bunz \etal \cite{bulletproofs}.  For each proof transcript, the extractor first runs the inner product extractor, to get a witness $\evaluatedleftpolynomial, \evaluatedrightpolynomial$ to the inner product argument such that $P = \hgenerator^{\mu}\vectorgenerator^{\evaluatedleftpolynomial}\vectorhgenerator^{\evaluatedrightpolynomial} \wedge \langle\evaluatedleftpolynomial, \evaluatedrightpolynomial\rangle = \evaluatedinnerproductpolynomial$. With this witness, and using two valid transcripts, one can compute linear combinations of (\ref{eq:p-equality}), we can compute $\alpha,\blindingblindings, \leftvector, \rightvector, \blindingvectorleft,\blindingvectorright$ such that $\commitmentvectors = \hgenerator^{\alpha}\vectorgenerator^{\leftvector}\vectorhgenerator^{\rightvector}$ and $\commitmentblinders = \hgenerator^{\blindingblindings}\vectorgenerator^{\blindingvectorleft}\vectorhgenerator^{\blindingvectorright}$. Such an extraction of these values proceeds as follows: 
\begin{align*}
& \commitmentvectors \cdot \commitmentblinders^{\challenge'} = \hgenerator^{\blindingannouncement'}\vectorgenerator^{\evaluatedleftpolynomial'}\vectorhgenerator^{\evaluatedrightpolynomial'}
& \commitmentvectors \cdot \commitmentblinders^{\challenge''} = \hgenerator^{\blindingannouncement''}\vectorgenerator^{\evaluatedleftpolynomial''}\vectorhgenerator^{\evaluatedrightpolynomial''}.
\end{align*}
Combining both relations we have 
\begin{align*}
\commitmentblinders = \left(\hgenerator^{\blindingannouncement' - \blindingannouncement''}\vectorgenerator^{\evaluatedleftpolynomial' - \evaluatedleftpolynomial''}\vectorhgenerator^{\evaluatedrightpolynomial' - \evaluatedrightpolynomial''}\right)^{\frac{1}{\challenge' - \challenge''}}.
\end{align*}
The extraction of $\commitmentvectors$ follows. 

Using these representations of $\commitmentvectors$ and $\commitmentblinders$, as well as $\evaluatedleftpolynomial^i$ and $\evaluatedrightpolynomial^i$ with $i\in\{', '', '''\}$, we have that 
\begin{align*}
\evaluatedleftpolynomial^i &= \leftvector + \blindingvectorleft\challenge^i \\
\evaluatedrightpolynomial^i &= \rightvector + \blindingvectorright\challenge^i.
\end{align*}
If these do not hold for all challenges, then the prover has found two distinct representations of the same group element, yielding a non-trivial discrete logarithm relation. 

Next, we extract the values $\commitmentcoefficients_i$ with $i\in\{1, 2\}$ from equation (\ref{eq:commitments-polys}) as follows: 
\begin{align*}
\generator^{\evaluatedinnerproductpolynomial'}\hgenerator^{\blindingevalippoly'} &=\commitmentinnerproduct\cdot\commitmentcoefficients_1^{\challenge'}\cdot\commitmentcoefficients_2^{\challenge'^2} \\
\generator^{\evaluatedinnerproductpolynomial''}\hgenerator^{\blindingevalippoly''} &=\commitmentinnerproduct\cdot\commitmentcoefficients_1^{\challenge''}\cdot\commitmentcoefficients_2^{\challenge''^2} \\
\generator^{\evaluatedinnerproductpolynomial'''}\hgenerator^{\blindingevalippoly'''} &=\commitmentinnerproduct\cdot\commitmentcoefficients_1^{\challenge'''}\cdot\commitmentcoefficients_2^{\challenge'''^2}
\end{align*}
which we can combine to get the following representation of $\commitmentcoefficients_2$:
\begin{align*}
\commitmentcoefficients_2 = \left( 
\left( 
\generator^{L}
\hgenerator^{R}
\right)^{\frac{1}{\challenge'- \challenge'''}}
\right)^{\frac{\challenge' + \challenge'''}{\challenge' + \challenge''}}
\end{align*}
with $L = \frac{\evaluatedinnerproductpolynomial'-\evaluatedinnerproductpolynomial''}{\challenge'- \challenge''} - 
\frac{\evaluatedinnerproductpolynomial' - \evaluatedinnerproductpolynomial'''}{\challenge'- \challenge'''}$ and $R = \frac{\blindingevalippoly'-\blindingevalippoly''}{\challenge'- \challenge''} - 
\frac{\blindingevalippoly' - \blindingevalippoly'''}{\challenge'- \challenge'''}$. Extractions of $\commitmentcoefficients_1$ and $\commitmentinnerproduct$ follow. 

If for any transcript, we have that 
\begin{equation*}
\evaluatedinnerproductpolynomial^i \neq \innerproduct + \coefficientip{1} \challenge^i + \coefficientip{2} \challenge^{i2}
\end{equation*}
with $i\in\{', '', '''\}$, then the extractor has again found a non trivial discrete logarithm relation. Let $P(X) = \langle\leftpolynomial(X), \rightpolynomial(X)\rangle$. Due to the validity of the transcripts, we have that $P(X)$ equals $t(X) = \innerproduct +\coefficientip{1} X + \coefficientip{2} X^2$ at least in the different challenges. In other words, the polynomial $P(X) - t(X)$ has at least three roots, and is of degree 2, hence it must be the zero polynomial. Therefore, we have that $t(X) = P(X)$. This implies that the zero coefficient of $t(X)$, namely $\innerproduct$, equals $\langle\leftvector, \rightvector\rangle$ and we have a valid witness for the statement. $\hfill\blacksquare$
\section{Formal Analysis of \innerProdName}
\label{sec:formal-analysis}
In this section we formally define the properties we expect out of \innerProdName, privacy and verifiability. We use
game based proofs to show that \innerProdName indeed provides these properties. We
formally define them here, and include the proofs subsequently. To model the experiments of privacy and verifiability, we define the following five functions:
\newcommand{\svmweights}{W}
\newcommand{\allproofs}{\Theta}
\newcommand{\evalvector}{\texttt{EvalSVM}}
\begin{itemize}
    \item  \setup($\secparam$): Which is defined exactly as in Procedure~\ref{proc:setup}. We omit the notation of the cryptographic material, and consider it implicit. We represent the set of parameters of the SVM model (normalisation mean, normalisation scale, SVM weight and SVM intercept) by $\svmweights$.
    \item \genproof($\sensorvector$): Generates the \innerProdName proof by running Procedure~\ref{proc:diff},~\ref{proc:sum},~\ref{proc:std}, and~\ref{proc:score}. Mainly, it runs all steps of the proof except for the submission step. We simplify the representation of the resulting tuple by $[\hashsensors, \allproofs, \openingscore, \randomresult]$, where $\allproofs$ consists of all intermediate proofs and commitments.
    \item \submitreq($[\hashsensors, \allproofs, \openingscore, \randomresult]$): Submits the output of \genproof($\sensorvector$) by sending it to the verifier (Procedure~\ref{proc:send}).
    \item \verifreq($[\hashsensors, \allproofs, \openingscore, \randomresult]$): Runs all procedures defined in the Verification phase of \innerProdName, mainly Procedures~\ref{proc:verif-diff}, \ref{proc:verif-sum}, \ref{proc:verif-std} and~\ref{proc:verif-score}. 
    \item \evalvector($\sensorvector$): Generates the result, $\openingscore$, corresponding to $\sensorvector$, as defined in the \innerProdName proof, but excluding the cryptographic mechanisms. 
\end{itemize}

\subsection{Privacy}
The goal of \innerProdName is that no information is leaked from the input vector 
other than the result of the SVM model. To model this, in Figure~\ref{fig:bpriv-game} we define an experiment, $\sprivacy{\adver}{\challenger}{\bit}$,
between an adversary, $\adver$, and a challenger, $\challenger$. The latter chooses
a bit $\bit\in\{0,1\}$, uniformly at random, which is given as input to the experiment.
The adversary controls
the \innerProdName verifier, and its goal is to distinguish the submissions of two different input vectors. 
The adversary is given access to an oracle, $\oraclesubmit()$, which takes
as input two vectors of size $\sizevector$, runs \genproof() over them, and submits a result. Note that it is the adversary who chooses the vectors over which the \innerProdName is executed and may modify the weights of the SVM model outputted by \setup(). Therefore, this experiment models the malicious choice of the SVM parameters, as well as any possible choice of input vector. 
Depending on the bit, $\bit$, the oracle 
submits the result of one vector or the other, by running \submitreq(). However, to avoid a trivial win 
by the adversary, 
the submitted SVM score is always computed over the first vector. Hence, the experiment
simulates the proof of the second vector by running $\simresult()$.
The adversary may call this oracle
as many times as it wishes. By the end of the experiment, the adversary outputs a bit, 
$\outputbit\in\{0,1\}$. The adversary wins if $\outputbit = \bit$ with non-negligible probability with respect to the security parameter, $\secparam$.

\begin{figure}[tbp]
  \centering
  {\small
  \begin{tabular}{@{}ll@{}}
    \toprule
    \multicolumn{2}{@{}l}{$\sprivacy{\adver}{\challenger}{\bit}(\secparam)$:} \\
    & \svmweights $\leftarrow \setup(\secparam)$ \\
    & $\outputbit \gets \adver^{\oraclesubmit}(\svmweights)$ \\
    & Output $\outputbit$ \\[1mm]
    \multicolumn{2}{@{}l}{$\oraclesubmit(v_1, v_2)$:} \\
    & Let $[\hashsensors^1, \allproofs^1, \openingscore^1, \randomresult^1]\leftarrow \genproof(v_1, \svmweights)$ \phantom{ }\\
    & Let $[\hashsensors^2, \allproofs^2, \openingscore^2, \randomresult^2]\leftarrow \genproof(v_2, \svmweights)$ \\
    & Let $\allproofs^2 \leftarrow \simresult(\hashsensors^2, \openingscore^1, \randomresult^1)$ \\
    & return $\submitreq([\hashsensors^b, \allproofs^b, \openingscore^1, \randomresult^1])$ \\
    \bottomrule
  \end{tabular}
  }
  \caption{In the privacy experiment
    $\sprivacy{\adver}{\challenger}{\bit}$, the adversary $\adver$ has access to
    the oracle $\oraclesubmit$ and controls zkSVM verifier.
  }
  \label{fig:bpriv-game}
\end{figure}

\begin{theorem}
There exists a \textup{$\simresult$} algorithm, such that no PPT adversary can win the~\innerProdName privacy experiment with colluding verifier with probability non-negligibly better than~1/2 with respect to $\secparam$. 
\end{theorem}

\noindent\textit{Proof.}
To prove that \innerProdName provides privacy we proceed by a series of games. We start with the adversary playing the privacy
experiment with $\bit = 0$, and after a sequence of game step transitions, the adversary finishes playing the ballot privacy experiment
with $\bit = 1$. We argue that each of these steps are indistinguishable, and therefore the results follows.
The proof proceeds along the following sequence of games:
\newcommand{\gameobject}{\textbf{G}}
\begin{description}
\item[Game $\gameobject_0$] Let game $\gameobject_0$ be the $\sprivacy{\adver}{\challenger}{0}(\secparam)$ game (see Figure~\ref{fig:bpriv-game}).
\item[Game $\gameobject_1$] Game $\gameobject_1$ is as in $\gameobject_0$, but now $\oraclesubmit$ always computes a simulation of $\allproofs$ regardless of the bit. Mainly, $\oraclesubmit$ proceeds as follows:

\vspace{3mm}
{\centering{
\small
\begin{tabular}{@{}ll@{}}
    \bottomrule
    \multicolumn{2}{@{}l}{$\oraclesubmit(v_1, v_2)$:} \\
    & Let $[\hashsensors^1, \allproofs^1, \openingscore^1, \randomresult^1]\leftarrow \genproof(v_1, \svmweights)$ \phantom{ }\\
    & Let $[\hashsensors^2, \allproofs^2, \openingscore^2, \randomresult^2]\leftarrow \genproof(v_2, \svmweights)$ \\
    & Let $\allproofs^1 \leftarrow \simresult(\hashsensors^1, \openingscore^1, \randomresult^1)$ \\
    & Let $\allproofs^2 \leftarrow \simresult(\hashsensors^2, \openingscore^1, \randomresult^1)$ \\
    & return $\submitreq([\hashsensors^1, \allproofs^1, \openingscore^1, \randomresult^1])$ \\
    \bottomrule
  \end{tabular}\\[3mm]
}}

The function $\simresult$ proceeds by simulating all zero knowledge proofs contained in $\allproofs$.
Because all these proofs are Zero Knowledge Proofs, and hence have the Special Honest-Verifier Zero-Knowledge property, there exists a simulation algorithm such that $\adver$ cannot distinguish between a real and a simulated proof. Note that at this point of the experiment, the commitment $\hashsensors$ and all commitments in $\allproofs$ correspond to those of $v_1$---only the zero knowledge proofs in $\allproofs$ are simulated. 
\item[Game $\gameobject_2$] Game $\gameobject_2$ is as in $\gameobject_1$, but now, instead of returning
\[\submitreq([\hashsensors^1, \allproofs^1, \openingscore^1, \randomresult^1]),\]
the oracle $\oraclesubmit$ returns 
\[\submitreq([\hashsensors^2, \allproofs^2, \openingscore^1, \randomresult^1]).\]
In $\gameobject_2$ the view of the adversary is identical to the one of $\sprivacy{\adver}{\challenger}{1}(\secparam)$. Only thing that remains is to prove that $\gameobject_1$ and $\gameobject_2$ are indistinguishable
\end{description}
Given the Special Honest-Verifier Zero-Knowledge property of the proofs, we know that the simulated view is random. Given that both simulations are equally distributed, it is infeasible to distinguish between $\allproofs^1$ and $\allproofs^2$. Similarly, given the perfectly hiding property of Pedersen commitments, no adversary can distinguish between $\hashsensors^1$ and $\hashsensors^2$. 
Clearly the resulting view is independent of $\bit$. And privacy follows.
$\hfill\blacksquare$

\subsection{Verifiability}
The other goal of \innerProdName is that an adversary cannot convince a verifier that the result is not linked to the committed vector as defined by the protocol. To model this, in Figure~\ref{fig:verif-game} we define an experiment, $\verifexp{\adver}{\challenger}$,
between an adversary, $\adver$, and a challenger, $\challenger$. Informally, verifiability 
ensures that a result that is not the outcome of the model evaluation over the committed vector cannot have a valid proof. During the experiment the adversary has access to the weights of the model, and generates an input vector, a result, and its corresponding proof material. The adversary wins the experiment if the result does not correspond the the SVM execution of the committed vector, and the verifier validates. We formally describe the experiment in Figure~\ref{fig:verif-game}.

\begin{figure}[t]
  \centering
  {\small
  \begin{tabular}{@{}ll@{}}
    \toprule
    \multicolumn{2}{@{}l}{$\verifexp{\adver}{\challenger}(\secparam)$:} \\
    & \svmweights $\leftarrow \setup(\secparam)$ \\
    & $[\sensorvector, \hashsensors, \allproofs, \openingscore, \randomresult] \gets \adver(\svmweights)$ \\
    & If $\evalvector(\sensorvector) = \openingscore$ return 0\\
    & If $\verifreq([\hashsensors, \allproofs, \openingscore, \randomresult]) = \bot$  \phantom{ }\\ 
    & then return 0, else return 1. \\[1mm]
    \bottomrule
  \end{tabular}
  }
  \caption{In the verifiability experiment
    $\verifexp{\bit}{\adver}$, the adversary $\adver$ needs to submit a result not corresponding to the committed vector.
  }
  \label{fig:verif-game}
\end{figure}

\begin{theorem}
No PPT adversary can win the \innerProdName verifiability experiment with non-negligible probability with respect to $\secparam$. 
\end{theorem}

\noindent\textit{Proof.}
At the end of the proof generation procedure, the prover (in our case the adversary) outputs a commitment, $\hashsensors$, a tuple of intermediate cryptographic material, $\allproofs$, and a score, $\openingscore$, together with the randomness associated to the commitment of the score, $\randomresult$. The result follows from the soundness property of ZKPs and the binding property of Pedersen commitments. Let us extend the cryptographic material associated with the vector. We have that 
$\allproofs = [\Delta, \text{M}, \text{M}', \Lambda, \Lambda']$, 
with
\begin{align*}
    \Delta = & \left[\diffcomm, \hashsensoriter, \proofequality, \proofzero\right], \\
    \text{M} = &\left[\avgcomm, \proofipavg\right], \\
    \text{M}' = &\left[\avgcommdiff, \proofipavgdiff\right], \\ 
    \Lambda = &\left[\gcommitmentaddition,\hcommitmentaddition,\hcommitmentsensor,\proofequalityg, \proofequalityh,\proofequalitys,\right. \\
    &\phantom{hello}\left.\commitmentvariance, \proofipvariance, \commitmentstd, \proofsqrtstd\right],\\
    \Lambda' = &\left[\gcommitmentadditiondiff,\hcommitmentadditiondiff,\hcommitmentsensordiff,\proofequalitygdiff, \proofequalityhdiff,  \right.\\
    & \left. \proofequalitysdiff,
\commitmentvariancediff, \proofipvariancediff, \commitmentstddiff, \proofsqrtstddiff\right].
\end{align*}
The proof verifies all proofs. In particular
\begin{itemize}
    \item Procedure~\ref{proc:verif-diff} first generates the iterated generators, and checks that indeed $\hashsensors$ and $ \hashsensoriter$ commit to the same opening, by verifying $\proofequality$. Then, it checks that $\diffcomm$ commits to the difference of $\hashsensors$ and $\hashsensoriter$, in all entries but the last, which contains a zero, by verifying $\proofzero$. 
    \item Procedure~\ref{proc:verif-sum} checks that $\avgcomm$ commits to the sum of the elements committed in $\hashsensors$ by verifying $\proofipavg$. It does the same for the difference vector.
    \item Procedure~\ref{proc:verif-std} first checks that $\gcommitmentaddition$ and $\hcommitmentaddition$ commit to the sum committed in $\avgcomm$ by verifying $\proofequalityg$ and $\proofequalityh$ respectively. Then, it checks that $\hcommitmentsensor$ commits to the same values as $\hashsensors$ by verifying $\proofequalitys$. With these verified commitments, the verifier can check that $\commitmentvariance$ commits to $N^2\langle \sensorvector - \vectoraverage, \sensorvector - \vectoraverage\rangle$, in other words, that it commits to a factor of the variance. It does so by running the algebraic operations of Equation~(\ref{eq:variance-relation}), and verifying $\proofipvariance$. Finally, to check that $\commitmentstd$ commits to the standard deviation of the vector committed in $\hashsensors$, it simply needs to check that it contains the square root of $\commitmentvariance$. It does so by verifying $\proofsqrtstd$. The same is performed for the difference vector. 
\end{itemize}
Given that all these proofs are Zero Knowledge Proofs, which provide the knowledge soundness property, the commitments of the SVM features, $\avgcomm,\avgcommdiff,\commitmentstd, \commitmentstddiff$, do not contain the expected features of the vector committed in $\hashsensors$ with negligible probability. It only remains to prove that the result indeed corresponds to the SVM function executed with these features as inputs. 
\begin{itemize}
    \item Procedure~\ref{proc:verif-score} first leverages the homomorphic properties of the commitment scheme. In this way, the verifier obtains a commitment of the linear combination of the values committed in $\avgcomm,\avgcommdiff,\commitmentstd, \commitmentstddiff$. This results in the commitment of the result, as described in Equation~(\ref{eq:linearcombinationscommitments}). This operation is conducted solely by the verifier, avoiding any possible attacks by the adversary. Finally, the verifier checks that the submitted score, and the corresponding opening key, indeed correspond to the locally computed commitment. Only if this is true, the verifier validates. 
\end{itemize}
Given that the commitment scheme used in \innerProdName is computationally binding, a PPT adversary has no more than negligible probability of submitting an opening that does not correspond to the committed value. 

Given that $\evalvector$ and the proof of \innerProdName compute the exact same operations over the input vector, the result that the adversary has no more than negligible probability of wining the verifiability experiment follows.
$\hfill\blacksquare$

\end{document}